\documentclass[conference]{IEEEtran}
\IEEEoverridecommandlockouts
\usepackage{cite}
\usepackage{amsmath,amssymb,amsfonts,amsthm}
\usepackage{algorithmic}
\usepackage{graphicx}
\usepackage{textcomp}
\usepackage{xcolor}

\usepackage{abbrevs}

\theoremstyle{plain}

\newtheorem{theorem}{Theorem}

\newcommand{\dif}{\mathrm{d}}

\newcommand{\lambdau}{\lambda^\mathrm{u}}

\newcommand{\lfa}{\lambda^\mathrm{FA}}

\newcommand{\pgfl}{\textsc{pgfl}\xspace}
\newcommand{\traj}{X} % Trajectory state
\newcommand{\stseq}{x} % State sequence
\newcommand{\tb}{\beta} % Trajectory birth time
\newcommand{\td}{\varepsilon} % Trajectory death time
\newcommand{\tlen}{\ell} % Trajectory length
\newcommand{\targetStateSpace}{\mathcal{X}} % Base state space for target state
\newcommand{\trajStateSpace}{\mathcal{T}} % Trajectory state space
\newcommand{\trackTable}{\mathbb{T}}

\newcommand{\trackori}{\textsc{to}\xspace}
\newcommand{\conv}[2]{\left\langle #1 ; #2 \right\rangle}
\newcommand{\convinline}[2]{\langle #1 ; #2 \rangle}

\newcommand{\infvec}{y}
\newcommand{\infmat}{Y}

\def\BibTeX{{\rm B\kern-.05em{\sc i\kern-.025em b}\kern-.08em
    T\kern-.1667em\lower.7ex\hbox{E}\kern-.125emX}}
    
\begin{document}

\title{Poisson multi-Bernoulli mixture trackers:\\ continuity through random finite sets of trajectories
%\thanks{Identify applicable funding agency here. If none, delete this.}
}

\author{\IEEEauthorblockN{Karl Granstr\"om, Lennart Svensson\\ Yuxuan Xia}
\IEEEauthorblockA{\textit{Dept. of Electrical Eng.} \\
\textit{Chalmers Univ. of Tech.}\\
Gothenburg, Sweden \\
{\footnotesize\texttt{firstname.lastname@chalmers.se}}}
\and
\IEEEauthorblockN{Jason Williams}
\IEEEauthorblockA{\textit{National Security and ISR Division} \\
\textit{Defence Science and Tech. Group}\\
Edinburgh, Australia \\
{\footnotesize\texttt{jason.williams@dst.defence.gov.au}}}
%\and
%\IEEEauthorblockN{Lennart Svensson}
%\IEEEauthorblockA{\textit{dept. name of organization (of Aff.)} \\
%\textit{name of organization (of Aff.)}\\
%City, Country \\
%email address}
\and
\IEEEauthorblockN{\'Angel F. Garc\'ia-Fern\'andez}
\IEEEauthorblockA{\textit{Dept. of Electrical Eng. and Electronics} \\
\textit{Univ. of Liverpool}\\
Liverpool, UK \\
{\footnotesize\texttt{angel.garcia-fernandez@liverpool.ac.uk}}}
%\and
%\IEEEauthorblockN{Yuxuan Xia}
%\IEEEauthorblockA{\textit{dept. name of organization (of Aff.)} \\
%\textit{name of organization (of Aff.)}\\
%City, Country \\
%email address}
}

\maketitle

\begin{abstract}
The Poisson multi-Bernoulli mixture (PMBM) is an unlabelled multi-target distribution for which the prediction and update are closed. It has a Poisson birth process, and new Bernoulli components are generated on each new measurement as a part of the Bayesian measurement update. The PMBM filter is similar to the multiple hypothesis tracker (MHT), but seemingly does not provide explicit continuity between time steps. This paper considers a recently developed formulation of the multi-target tracking problem as a random finite set (RFS) of trajectories, and derives two trajectory RFS filters, called PMBM trackers. The PMBM trackers efficiently estimate the set of trajectories, and share hypothesis structure with the PMBM filter. By showing that the prediction and update in the PMBM filter can be viewed as an efficient method for calculating the time marginals of the RFS of trajectories, continuity in the same sense as MHT is established for the PMBM filter.
\end{abstract}


%%%%%%%%%%%%%%%%%%%%%%%%%%%%%%%%%%
%%%%%%%%%%%%%%%%%%%%%%%%%%%%%%%%%%
%%%%%%%%%%%%%%%%%%%%%%%%%%%%%%%%%%
%%%%%%%%%%%%%%%%%%%%%%%%%%%%%%%%%%
%%%%%%%%%%%%%%%%%%%%%%%%%%%%%%%%%%

\section{Introduction}
\label{sec:Introduction}
Multi-target tracking (\mtt) is a challenging problem due to the unknown correspondence of measurements and targets, referred to as data association. %\footnote{We consider problems with point measurements (as opposed to image measurements, or track before detect), where a set of measurements is received at each time, and the origin of each measurement is unknown.} 
Each new measurement received could be the continuation of some previously detected target, the first detection of a new target, or a false alarm. 

The multiple hypothesis tracker (\mht) introduced in \cite{Rei79} is a key method for addressing problems of this type. The algorithm maintains a series of global hypotheses for each possible measurement-target correspondence as measurements are received, along with a conditional state distribution for each target under each hypothesis. Each measurement could be hypothesised to represent the first detection of a new target, where the number of newly detected targets was given a Poisson distribution in order to provide a Bayesian prior. 

In \cite{MorCho86}, the model was made rigorous through random finite sequences, under the assumption that the number of targets present is constant but unknown, with a Poisson prior. Target state sequences were formed under each global hypothesis, and the Poisson distribution of targets remaining to be detected provided a Bayes prior for events involving newly detected targets. Hypotheses were constructed as being data-to-data, since no a priori data was assumed on target identity. This formulation was revisited in \cite{MorCho16}, and compared to random finite sets, and finite point processes.

The framework of random finite sets (\rfss) was developed to provide a systematic methodology for dealing with problems involving an uncertain number of targets \cite{Mah07}. In \cite{Wil12}, the Bayes filter for Poisson birth models was derived using \rfss, obtaining a result somewhat similar to \cite{MorCho86},  involving a Poisson distribution representing targets which are hypothesised to exist but remain to be detected, and a multi-Bernoulli mixture (\textsc{mbm}) representing targets that have been detected at some stage. Adopting the common modelling assumptions employed in \rfs (see, e.g., \cite{Mah03}), target appearance and disappearance was modelled.  The resulting, so called, \pmbm filter, did not formally establish track continuity, although a hypothesis structure similar to \mht has been observed \cite{BreChi17,BreChi18}. 

\mht was extended in \cite{CorCar13,CorCar14} to address a time-varying number of targets, incorporating birth of targets which are not immediately detected, necessitating equivalence classes of indistinguishable hypotheses. In comparison, in \cite{Wil12} the hypotheses are purely data-to-data assignments, the hypothesis space explicitly includes targets that remain to be detected, and target death subsequent to the final detection.

The labelled \rfs formulation of \cite{VoVo13} addresses the lack of continuity of unlabelled \rfss by incorporating a label (uniqueness of which is ensured through the model) into the target state; continuity is maintained by connecting estimates with the same label at different times. PGFL expressions for the joint target-measurement density over a time interval were given in \cite{Str17}, including modelling of target birth, target death, and target spawning using branching and immigration processes.  
%Track continuity could also be achieved using interval/smoothing filters; in \cite{Str17} \pgfl expressions for multiple target interval filters were provided with target birth based on both branching and immigration processes. 
A solution was proposed using saddle point approximation, however, a practical implementation was not provided in \cite{Str17}. In \cite{DelFru18}, a framework for performing inference on problems involving both distinguishable (previously detected) and indistinguishable (never detected) targets, was developed and applied to space situation awareness. Targets were characterized by sequences of measurements, under the non-standard assumption that the targets are \emph{``immediately detected upon entering the surveillance scene''} \cite[A.5]{DelFru18}.

A formulation of the target tracking problem as an unlabelled \rfs of trajectories was provided recently in \cite{SvenssonM:2014,GarSve16}. Within this set of trajectories framework, the goal of Bayesian \mtt is to compute the posterior density over the set of trajectories. One approach to do this is to let the multi-target state be the set of trajectories, and compute the posterior density recursively. Assuming standard point target motion and measurement models, in this paper we formulate two different set of trajectory problems: one in which the set of current (i.e., remaining) trajectories is tracked, and one in which the set of all trajectories up to the current time is tracked. For the two problem formulations, we assume multi-target densities of the \pmbm form and derive the \pmbm predictions and the \pmbm update. This results in two different \pmbm  set of trajectories filters: we call these tracking algorithms \emph{PMBM trackers}, to distinguish them from the \pmbm filter \cite{Wil12}, which is for sets of target states.

%%%%%%%%%%%%%%%%%%%%%

%{\color{red}In \cite[Sec. 2]{CorCar14} it is claimed that the set of trajectories density is \emph{``extremely challenging to compute''} and instead they seek to estimate the maximum a posteriori estimat -- in this paper we show that for standard modelling assumptions, this claim is not true: the density can indeed be computed. Further, in \cite[Sec. 3]{CorCar14} they claim that the MAP is intractable to compute, and give a solution conditioned on a discrete state variable that identifies all birth, death, and measurement-association events up to the current time. Again, we show that this is not necessary?
%
%Not detected split into two: unnoticed (eventually detected) and ghost (never detected). Expansion into hypotheses seems to be conditioned on \emph{``certain existence''}, which is less efficient than the Bernoulli with probability of existence.
%
%The example is a 1-D process.
%
%In \pmbm trackers, we do not need any indistinguishable hypotheses.}
%
%{\color{blue} In \cite[Following eq 2]{DelFru18} it seems  there needs to be a consequtive stream of measurements? Importantly, they assume that targets are detected immediately upon entering the scene.}
%
%{\color{green}In this paper, we show that it is possible to compute the set of trajectories density, and we do not require an assumption that targets are detected immediately upon entering the surveillance space.}

The prediction and update of \pmbm trackers are the main contribution in the paper. In addition to this, we discuss the \pmbm trackers in relation to the \pmbm filter \cite{Wil12}, the \mht \cite{Rei79,MorCho86}, and the \dglmb filter \cite{VoVo13}. Interestingly, the relations between the \pmbm trackers and the \pmbm filter show that the \pmbm filter implicitly contains trajectory information. We also show how the \pmbm trackers can be implemented for linear Gaussian models, and present results from a simulation study where we compare the results to the \dglmb filter \cite{VoVo13}.

Importantly, our paper shows that it is possible to compute the density over the set of trajectories even for complicated problems; in \cite[Sec. 2]{CorCar14} computing the set of trajectories density was thought to be \emph{``extremely challenging\ldots even for small problems''}. Further, we compute the density without requiring the assumption that targets are detected with probability one upon entering the scene.

\section{Problem formulation}

To clearly differentiate between a target state at a single time and a sequence of target states, we let \emph{target} denote the state at some time, and we let \emph{trajectory} denote a sequence of states. Thus, \emph{``set of targets''} and \emph{``target \rfs''} refers to formulations involving \rfss of target states at a single time (e.g., the unlabelled \rfs formulation in \cite{Wil12}), and \emph{``set of trajectories''} and \emph{``trajectory \rfs''} refers to formulations involving a \rfs of trajectories, or sequences of states.

Let $x_{k}$ denote a target state at time $k$, and let $z_{k}$ denote a measurement at time $k$. We utilise the standard multi-target dynamics model, defined in Assumption \ref{ass:Dynamics}, and the standard point target measurement model, defined in Assumption \ref{ass:Measurement}.

\begin{assumption}
	The multiple target state evolves according to the following time dynamics process: targets arrive at each time according to a non-homogeneous Poisson Point Process (PPP) with birth intensity $\lambda^\mathrm{b}(x_k)$, independent of existing targets; targets depart according to iid Markovian processes; the survival probability in state $x_k$ is $P^{\rm S}(x_k)$; target motion follows iid Markovian processes; the single-target transition density is $\pi^{x}(x_k|x_{k-1})$.%
	%
%	\begin{itemize}
%		\item Targets arrive at each time according to a non-homogeneous PPP with birth intensity $\lambda^\mathrm{b}(x_k)$, independent of existing targets
%		\item Targets depart according to iid Markovian processes; the survival probability in state $x_k$ is $P^{\rm S}(x_k)$
%		\item Target motion follows iid Markovian processes; the single-target transition PDF is $\pi^{x}(x_k|x_{k-1})$
%	\end{itemize}
	\label{ass:Dynamics}%
\end{assumption}%
\begin{assumption}%
	The multiple target measurement process is as follows: each target may give rise to at most one measurement; probability of detection in state $x_k$ is $P^{\rm D}(x_k)$; each measurement is the result of at most one target; false alarm measurements arrive according to a non-homogeneous PPP with intensity $\lambda^\mathrm{FA}(z_k)$, independent of targets and target-related measurements; each target-derived measurement is independent of all other targets and measurements conditioned on its corresponding target; the single target measurement likelihood is $f(z_k|x_k)$.%
	%
%	\begin{itemize}
%		\item Each target may give rise to at most one measurement; probability of detection in state $x_k$ is $\Pdet(x_k)$
%		\item Each measurement is the result of at most one target
%		\item False alarm measurements arrive according to a non-homogeneous PPP with intensity $\lambda^\mathrm{FA}(z_k)$, independent of targets and target-related measurements
%		\item Each target-derived measurement is independent of all other targets and measurements conditioned on its corresponding target; the single target measurement likelihood is $f(z_k|x_k)$
%	\end{itemize}%
	\label{ass:Measurement}%
\end{assumption}%
%The overall objective is to use the above models, together with a sequence of measurements, to obtain information about the targets. In this context, we distinguish between multi-target filtering (\mtf) and multi-target tracking (\mtt) as follows: the objective of \mtf is to estimate the number of targets, and to estimate their respective states; the objective of \mtt is to estimate the number of targets, and to estimate their respective sequences of states, i.e., the \emph{trajectories}.\footnote{In \cite{GarSve16}, this is referred to as \emph{multi-trajectory filtering}.}
%The focus of this paper is on \mtt. 
There are many ways in which a Multiple Target Tracking (\mtt) problem can be formulated. In this paper, we focus on the following two variants:
\begin{enumerate}
	\item \emph{The set of current trajectories:} the objective is to estimate the trajectories of the targets who are present in the surveillance area at the current time.
	\item \emph{The set of all trajectories:} the objective is to estimate the trajectories of all targets that have passed through the surveillance area at some point between time $0$ and the current time, i.e., both the targets who are present in the surveillance area at the current time, and the targets that have left the surveillance area (but were in the surveillance area at at least one previous time).
\end{enumerate}
Note that both problem formulations are equally valid; which one is relevant depends on the application. Indeed, in some applications, a solution with only current states (no trajectories) is the objective.

%%%%%%%%%%%%%%%%%%%%%%%%%%%%%%%%%%
%%%%%%%%%%%%%%%%%%%%%%%%%%%%%%%%%%
%%%%%%%%%%%%%%%%%%%%%%%%%%%%%%%%%%
%%%%%%%%%%%%%%%%%%%%%%%%%%%%%%%%%%
%%%%%%%%%%%%%%%%%%%%%%%%%%%%%%%%%%

\section{Random finite sets of trajectories}
\label{sec:RFS}

Compact representations of the multiple trajectory density for the standard point target models, cf. Assumptions~\ref{ass:Dynamics} and \ref{ass:Measurement}, are obtained using sets of trajectories, developed in the \rfs formalism in \cite{GarSve16}. %We extend the derivation of the target \pmbm filter in \cite{Wil12} to a trajectory \pmbm.
Expressions for integration, Bayes prediction and Bayes update may be found there. In this section, we first review the trajectory state representation, and then present densities and probability generating functionals for sets of trajectories.

\subsection{Single trajectory}
\label{sec:SingleTrajectoryStateRepresentation}
Let $\targetStateSpace$ represent the base state space, e.g., $\targetStateSpace=\mathbb{R}^4$, representing position and velocity in two dimensions. We use the trajectory state model proposed in \cite{SvenssonM:2014,GarSve16}, in which the trajectory state is a tuple
\begin{align}
	\traj = \left(\tb,\td,\stseq_{\tb:\td}\right)
\end{align}
where $\tb$ is the discrete time of the trajectory birth, i.e., the time the trajectory begins; $\td$ is the discrete time of the trajectory's most recent state, i.e., the time the trajectory ends. If $k$ is the current time, $\td<k$ means that the trajectory ended at time $\td$, and $\td=k$ means that the trajectory is ongoing; $\stseq_{\tb:\td}$ is, given $\tb$ and $\td$, the sequence of states 
		%\begin{subequations}
		\begin{align}
			& x_{\tb},x_{\tb+1},\ldots,x_{\td-1},x_{\td},
		\end{align}
		where $x_{k}\in\targetStateSpace$ for all $k\in\{\tb,\ldots,\td\}$.
		%\end{subequations}
%\begin{itemize}
%	\item $\tb$ is the discrete time of the trajectory birth, i.e., the time the trajectory begins.
%	\item $\td$ is the discrete time of the trajectory's most recent state, i.e., the time the trajectory ends. If $k$ is the current time, $\td<k$ means that the trajectory ended at time $\td$, and $\td=k$ means that the trajectory is ongoing.
%	\item $\stseq_{\tb:\td}$ is, given $\tb$ and $\td$, the sequence of states 
%		\begin{subequations}
%		\begin{align}
%			& x_{\tb},x_{\tb+1},\ldots,x_{\td-1},x_{\td}, \\
%			& x_{k}\in\targetStateSpace,\forall k\in\{\tb,\ldots,\td\}.
%		\end{align}
%		\end{subequations}
%\end{itemize}
This gives a trajectory of length $\tlen = \td-\tb+1$ time steps. The trajectory state space at time $k$ is \cite{GarSve16}
\begin{align}
	& \trajStateSpace_{k} = \uplus_{(\tb,\td)\in I_{k}} \{\tb\}\times\{\td\}\times\targetStateSpace^{\td-\tb+1},
\end{align}
where $\uplus$ denotes disjoint set union, $I_{k} = \{ (\tb,\td) : 0\leq \tb \leq \td \leq k \}$ and $\targetStateSpace^{\tlen}$ denotes $\tlen$ Cartesian products of $\targetStateSpace$. The trajectory state density factorises as follows\footnote{``$p_{k|k'}$'' is the density of $\traj_k$ given measurements up to and including time $k'\leq k$.}
\begin{align}
	p_{k|k'}(\traj) = p_{k|k'}(\stseq_{\tb:\td} | \tb,\td) P_{k|k'}(\tb,\td), \label{eq:trajectory_state_density}
\end{align}
where, if $\td<\tb$, then $P_{k|k'}(\tb,\td)$ is zero. Integration is performed as follows \cite{GarSve16},
{
\begin{subequations}
\begin{align}
	&\int p(\traj) \diff \traj =   \iiint p(\stseq_{\tb:\td} | \tb,\td) P(\tb,\td) \diff \stseq_{\tb:\td} \diff\td \diff\tb \\
	%= & \sum_{\tb,\td} \left[ \int p(\stseq_{\tb:\td} | \tb,\td) \diff \stseq_{\tb:\td} \right] P(\tb,\td), \\
	&=  \sum_{\tb,\td} \left[ \idotsint p(x_{\tb},\ldots,x_{\td} | \tb,\td) \diff x_{\tb}\ldots \diff x_{\td} \right] P(\tb,\td) .
	%\int p(\stseq_{\tb:\td} | \tb,\td) & \diff \stseq_{\tb:\td}  \nonumber \\
	%= & \idotsint p(x_{\tb},\ldots,x_{\td} | \tb,\td) \diff x_{\tb}\ldots \diff x_{\td}
\end{align}%
\end{subequations}
}%
\subsection{Multiple trajectories}
We now move to the case of a \emph{set} of trajectories. Analogously to a set of targets, a set of trajectories is denoted as $\setX_k \in \mathcal{F}(\trajStateSpace_{k})$, where $\mathcal{F}(\trajStateSpace_{k})$ is the set of all finite subsets of $\trajStateSpace_{k}$.
%Recalling the notation $\traj_k = [\setX_1;\dots;\setX_{k}]\in\trajStateSpace_T$, integration over the trajectory is defined as:
%%
%%
%\begin{equation}
%\int{g(\traj_k) \dif \traj_k} = \idotsint{g([\setX_1;\dots;\setX_{k}]) \delta \setX_1 \cdots\delta \setX_{k}}.
%\end{equation}
%
%Subsequently, 
Let $g(\setX_k)$ be a function on a set of trajectories $\setX_k$. Integration over sets of trajectories is defined as regular set integration:
\begin{multline}
\int{g(\setX_{k}) \delta \setX_{k}}
\triangleq g(\emptyset) + \\ 
\sum_{n=1}^{\infty}\frac{1}{n!}\idotsint{g(\{\traj^1_k,\dots,\traj^n_k\})\dif \traj^1_k \cdots \dif \traj^n_k} . \label{eq:SetTrajIntegral}
\end{multline}
%
%One subtlety in this expression is that multiple targets could appear at the discrete point $\traj_k=[\emptyset;\dots;\emptyset]$ with non-zero measure; this cannot be accommodated in the standard formulation of \rfss as a simple point process. While this could be addressed by using the extension of \rfss to multisets in \cite{???} \textit{(Streit? Vo?)}, since the point represents a target that did not exist at any time, we instead craft our birth process to only permit introduction of targets that \textit{do exist} at some time, so that the measure of this discrete point is identically zero. 

The set of trajectories density function $f_{\setX}(\setX)$ is defined analogously to the set of targets density function. The probability generating functional (\pgfl), see, e.g., \cite{Mah07}, is a useful tool for manipulating and understanding \rfs densities. The \pgfl for a trajectory \rfs density is defined like the \pgfl for a target \rfs density: using a test function $h(\traj) : \trajStateSpace \rightarrow \mathbb{R}$, the \pgfl is
%(where $\trajStateSpace$ is the space $T$-dimensional tuples of Bernoulli sets of $\targetStateSpace$), 
%
\begin{align}
G_\setX[h] &\triangleq \int{h^\setX f_\setX(\setX) \delta \setX} \\
&\triangleq f_\setX(\emptyset) + \sum_{n=1}^{\infty}\frac{1}{n!}\idotsint{\left[\prod_{i=1}^n h(\traj^i)\right]} \notag\\
&\quad \times f_\setX(\{\traj^1,\dots,\traj^n\})\dif \traj^1 \cdots \dif \traj^n , 
\label{eq:TrajPGFL}
\end{align}
where $h^{\setX}$ is set power, defined as $h^{\setX}= 1$ if $\setX=\emptyset$ and $h^{\setX}=\prod_{\traj\in\setX} h(\traj)$ if $\setX\neq\emptyset$.
%Thus, while the state space $\trajStateSpace$ is non-standard in point target \mtt, in which the target state space

%We denote by $\omega(\traj_k)=x_{\td}$ the last state in the trajectory, and by $\bar{\omega}(\traj_k)=[x_{\tb};...;x_{\td-1}]$ the trajectory excluding the last state, so that $\stseq_{\tb:\td}=[\bar{\omega}(\traj_k);\omega(\traj_k)]$. We denote by $\psi(\traj_k)=\omega(\bar{\omega}(\traj_k))=x_{\td-1}$ the second-last Bernoulli state in the trajectory. 
%We define the delta function on the trajectory state space to be such that:
%%
%\begin{equation}
%\int{\delta_{\traj'}(\traj)f(\traj)\dif \traj} \triangleq f(\traj')
%\end{equation}

%We denote by $\omega(\traj_k)=\setX_{k}$ the last Bernoulli state in the trajectory, and by $\bar{\omega}(\traj_k)=[\setX_1;...;\setX_{k-1}]$ the trajectory excluding the last state, so that $\traj_k=[\bar{\omega}(\traj_k);\omega(\traj_k)]$. We denote by $\psi(\traj_k)=\omega(\bar{\omega}(\traj_k))=\setX_{k-1}$ the second-last Bernoulli state in the trajectory. We define the delta function on the hybrid space to be such that:
%%
%\begin{equation}
%\int{\delta_{y'}(y)f(y)\dif y} \triangleq f(y')
%\end{equation}

A trajectory \ppp is analogous to a target \ppp, and has set density and \pgfl
\begin{subequations}
	\begin{align}
		f^\mathrm{ppp}(\setX) & = e^{-\int \lambda(\traj)\diff\traj} \lambda^{\setX} \\ % \prod_{\traj\in\setX} \lambda(\traj), \\
		G^\mathrm{ppp}[h] &= \exp\left\{\conv{\lambda}{h-1}\right\},
	\end{align}
\end{subequations}
where $\conv{f}{g} = \int f(\traj) g(\traj) \diff \traj$. The trajectory \ppp intensity $\lambda(\cdot)$ is defined on the trajectory state space $\trajStateSpace_{k}$, i.e., realisations of the \ppp are trajectories with a birth time, a time of the most recent state, and a state sequence.

A trajectory Bernoulli process is analogous to a target Bernoulli process, and has set density and \pgfl
\begin{subequations}
\begin{align}
f^\mathrm{ber}(\setX) &= \begin{cases}
1-r, & \setX = \emptyset \\
r f(\traj), & \setX = \{\traj\} \\
0, & \mbox{otherwise}
\end{cases}  \\
G^\mathrm{ber}[h] %&= 1 - r_{k|k'}^{i,a^i} + r_{k|k'}^{i,a^i} \int{ f_{k|k'}^{i,a^i}(\traj_k)h(\traj_k)\dif \traj_k}  \\
& = 1 - r + r \conv{f}{h}
\end{align}%
\label{eq:Bernoulli}%
\end{subequations}%
Here, $f(\cdot)$ is a trajectory state density \eqref{eq:trajectory_state_density}, and $r$ is the Bernoulli probability of existence. Together, $r$ and $f(\cdot)$ can be used to find the probability that the target trajectory existed at a specific time, or find the probability that the target state was in a certain area at a certain time. Trajectory \mb \rfs and trajectory \mbm \rfs are both defined analogously to target \mb \rfs and target \mbm \rfs: a trajectory \mb is the disjoint union of a multiple trajectory Bernoulli \rfs; trajectory \mbm \rfs is an \rfs whose density is a mixture of trajectory \mb densities.

\section{PMBM trackers}
As in \cite{Wil12}, we hypothesise a multi-target conjugate prior\footnote{\emph{Multi-target conjugate prior} was defined in \cite{VoVo13} as meaning that \emph{"{\ldots}if we start with the proposed conjugate initial prior, then all subsequent predicted and posterior distributions have the same form as the initial prior."}} of the following Poisson Multi-Bernoulli Mixture (\pmbm) form, and observe that the form is preserved through prediction and update. The \pmbm \pgfl is
%
%\begin{subequations}
\begin{align}
&G_{k|k'}[h] = \exp\left\{\conv{\lambdau_{k|k'}}{h(\traj_k)-1}\right\} \label{eq:SetCPUnion} \\
&\times \sum_{a\in{\cal A}^{k|k'}} w_{k|k'}^{a} \prod_{i\in{\trackTable}_{k|k'}}\left( 1 - r_{k|k'}^{i,a^i} + r_{k|k'}^{i,a^i} \conv{f_{k|k'}^{i,a^i}}{h} \right) . \nonumber
\end{align}
%\end{subequations}
%
This \pgfl form states that the set of trajectories $\setX_k$ is an independent union of a \ppp with intensity $\lambdau_{k|k'}$, and a multi-Bernoulli mixture (\mbm) with Bernoulli parameters $r_{k|k'}^{i,a^i}$ and $f_{k|k'}^{i,a^i}(\cdot)$. 
%The \pgfl $G^\mathrm{ppp}_{k|k'}[h]$ represents the \ppp distribution:
%\begin{subequations}
%\begin{align}
%	f_{k|k'}(\setX) & = e^{-\int \lambdau_{k|k'}(\traj)\diff\traj} \prod_{\traj\in\setX} \lambdau_{k|k'}(\traj), \\
%	G^\mathrm{ppp}_{k|k'}[h] &= \exp\left\{\conv{\lambdau_{k|k'}}{h(\traj_k)-1}\right\}. \label{eq:SetCPPPP}
%\end{align}
%\end{subequations}
%where $\conv{f}{g} = \int f(x) g(x) \diff x$. The \pgfl $G_{k|k'}^{i,a^i}[h]$ represents the Bernoulli distribution:
%%
%\begin{subequations}
%\begin{align}
%f_{k|k'}^{i,a^i}(\setX) &= \begin{cases}
%1-r_{k|k'}^{i,a^i}, & \setX = \emptyset \\
%r_{k|k'}^{i,a^i} f_{k|k'}^{i,a^i}(\traj_k), & \setX = \{\traj_k\} \\
%0, & \mbox{otherwise}
%\end{cases}  \\
%G_{k|k'}^{i,a^i}[h] %&= 1 - r_{k|k'}^{i,a^i} + r_{k|k'}^{i,a^i} \int{ f_{k|k'}^{i,a^i}(\traj_k)h(\traj_k)\dif \traj_k}  \\
%& = 1 - r_{k|k'}^{i,a^i} + r_{k|k'}^{i,a^i} \conv{f_{k|k'}^{i,a^i}}{h}
%\end{align}
%\label{eq:Bernoulli}
%\end{subequations}

The structure of the trajectory \pmbm \eqref{eq:SetCPUnion} is the same as the structure of the target \pmbm from \cite{Wil12}. The \ppp represents trajectories that are hypothesised to exist, but have never been detected, i.e., no measurement has been associated to them. We present \emph{``track oriented''} (\trackori) \pmbm trackers, where a track is initiated for each measurement. In the \mbm in \eqref{eq:SetCPUnion}, ${\trackTable}_{k|k'}$ is a track table with $n_{k|k'}$ tracks,  $a\in{\cal A}^{k|k'}$ is a possible global data association hypothesis, and for each global hypothesis $a$ and each track $i\in{\trackTable}_{k|k'}$, $a^i$ indicates which track hypothesis is used in the global hypothesis. For each track, there are $n_{k|k'}^{i}$ single trajectory hypotheses. The weight of the global hypothesis $a$ is $w_{k|k'}^{a} \propto \prod_{i\in{\trackTable}_{k|k'}} w_{k|k'}^{i,a^i}$, where $w_{k|k'}^{i,a^i}$ is the weight of single trajectory hypothesis $a^i$ from track $i$. The set of all measurement indices up to time $k$ is denoted ${\cal M}^k$, and ${\cal M}^{k}(i,a^i)$ is the history of measurements that are hypothesised to belong to hypothesis $a^i$ from track $i$. Due to page length constraints, further elaboration is omitted; please refer to \cite{Wil12} for details and discussion.

Depending on the problem formulation, the multi-target state $\setX_k$ that we are interested in is different. For the set of current trajectories, $\setX_k$ is the set of trajectories for which $0\leq\tb\leq\td=k$. For the set of all trajectories, $\setX_{k}$ is the set of trajectories for which $0\leq\tb\leq\td\leq k$. A \pmbm process is defined by the parameters of the {density}/{\pgfl},
\begin{align}
	\lambdau_{k|k'}, \left\{ w_{k|k'}^{i,a^i} , r_{k|k'}^{i,a^i} , f_{k|k'}^{i,a^i} \right\}
\end{align}
In the following subsections we will show how the parameters are predicted and updated, in order to track either the set of current trajectories, or the set of all trajectories.

\subsection{Prediction step}
In this section we describe the time evolution of the set of trajectories. A standard \ppp birth model is used, i.e., target birth at time step $k$ is modelled by a \ppp, with intensity
\begin{subequations}
\begin{align}
	\lambda_{k}^{{\rm B}}(\traj) &= \lambda_{k}^{{\rm B},\stseq}(\stseq_{\tb:\td} | \tb,\td) \Delta_{k}(\td) \Delta_{k}(\tb), \\
	\lambda_{k}^{{\rm B},\stseq}(\stseq_{k:k} | k,k) & = \lambda_{k}^{\mathrm{b}}(x_k),
\end{align}%
\label{eq:PoissonBirthIntensity}%
\end{subequations}%
where $\Delta(\cdot)$ denotes Kronecker's delta function. Note that it is possible to have alternative birth models, such as \mb birth, or \mbm birth. For those cases, the spatial densities of the Bernoulli birth components would be of the same form as in \eqref{eq:PoissonBirthIntensity}, i.e., for birth at time $k$ we have $\tb=\td=k$.

For the transition of existing targets, there is one alternative for each of the two problem formulations. %are two different alternatives, which correspond to the two different problem formulations. 
For the set of current trajectories, $P^{\rm S}(\cdot)$ is used in a way that is typical for tracking a set of targets, see, e.g., \cite{Wil12,VoVo13}. For the set of all trajectories, $P^{\rm S}(\cdot)$ is used as in \cite{GarSve16}. The trajectory state dependent probability of survival at time $k$ is defined as
\begin{align}
	P_{k}^{\rm S}(\traj) = P^{\rm S}(x_{\td})\Delta_{k}(\td). % \left\{\begin{array}{lcl} 0 & \text{if} &  \td \neq k \\ P^{\rm S}(x_{\td}) & & \td=k \end{array} \right. .
	\label{eq:trajectory_dependent_probability_of_survival}
\end{align}
  %ways in which we can use the probability of survival $P^{\rm S}(x)$ within set of trajectories tracking. 
%One corresponds directly to the way in which the probability of surivival is typically used in set of targets tracking, see, e.g., \cite{Wil12,VoVo13}; this leads to a tracker that tracks the current set of trajectories, i.e., the set of trajectories for which the last time step $\td$ is equal to the current time step $k$. The second way to use the probability of survival was used previously in \cite{GarSve16}; it leads to a tracker that tracks the set of all trajectories, i.e., both trajectories for which $\td=k$ and trajectories for which $\td<k$. 
%A few things are worth pointing out:
%\begin{itemize}
%	\item The birth time $\tb$ is necessarily equal to the current time step $k$, because otherwise ``a target born at time $k$'' would lose its meaning.
%	\item The death time $\td$ is also equal to the current time step $k$, because it cannot be less than $\tb$ and it does not make sense to give birth to an object whose trajectory extends into the future.
%	\item The state intensity $\lambda_{k}^\mathrm{B}(x_k)$ can be multi-modal, represented by a density mixture, e.g., with modes for the different birth locations that are likely.
%\end{itemize}
\subsubsection{Transition model for the set of current trajectories}
The Bernoulli \rfs transition density without birth is
{
\begin{subequations}
\begin{align}
	& f_{k|k-1}^{\rm c}(\setX | \setX') = \label{eq:set_of_current_trajectories_transition_density}\\
	& \left\{ \begin{array}{lcl}
			1, & & \setX'=\emptyset, \setX=\emptyset \\
			%0 & & \setX'=\emptyset, \setX\neq\emptyset \\
			1-P_{k-1}^{\rm S}(\traj'), & \text{if} & \setX'=\{\traj'\},\setX=\emptyset \\
			P_{k-1}^{\rm S}(\traj')\pi^{\rm c}(\traj|\traj'), & & \setX'=\{\traj'\},\setX=\{\traj\} \\
			0, & & \text{otherwise}
		\end{array} \right. \nonumber \\
	%& \pi^{\rm c}(\traj | \traj') =  \pi^{x}_{}(x_{\td} | x'_{\td'}) \delta_{\stseq'_{\tb':\td'}}(\stseq_{\tb:\td-1})  \Delta_{\td'+1}(\td)  \Delta_{\tb'}(\tb), %\\
	& \pi^{\rm c}(\traj | \traj') =  \pi_{}^{{\rm c},\stseq}(\stseq_{\tb:\td} | \tb,\td,\traj')   \Delta_{\td'+1}(\td)  \Delta_{\tb'}(\tb), \\
	& \pi^{{\rm c},\stseq}_{}(\stseq_{\tb:\td} | \tb,\td,\traj') = \pi^{x}_{}(x_{\td} | x'_{\td'}) \delta_{\stseq'_{\tb':\td'}}(\stseq_{\tb:\td-1}) ,
\end{align}%
\label{eq:set_of_current_trajectories_state_transition_density}%
\end{subequations}%
}%
where $\delta(\cdot)$ denotes Dirac's delta function. %, and the trajectory dependent probability of survival is defined in \eqref{eq:trajectory_dependent_probability_of_survival}. 
In this model, $P^{\rm S}(\cdot)$ is used as follows. If the target disappears, or ``dies'' (case $\setX'=\{\traj'\},\setX=\emptyset$ in \eqref{eq:set_of_current_trajectories_transition_density}), then the entire trajectory will no longer be a member of the set of current trajectories. If the target survives, then the trajectory is extended by one time step. The \pgfl of the multiple object transition density is
\begin{align}
	G_{k|k-1}^{\rm c}[h | \setX] = e^{\conv{\lambda_{k}^{\rm B}}{h-1}} \left(1-P_{k-1}^{\rm S} + P_{k-1}^{\rm S}\conv{\pi^{\rm c}}{h}\right)^{\setX}
	\label{eq:set_of_current_trajectories_state_transition_pgfl}
\end{align}
%Using this first model of the time evolution of the set of trajectories, 
The resulting prediction step is given in the theorem below.
\begin{theorem}\label{th:PredictionCurrentTrajectories}
	Assume that the distribution from the previous time step $G_{k-1|k-1}[h]$ is the \pgfl of the form given in (\ref{eq:SetCPUnion}), and that the transition model is of the kind (\ref{eq:set_of_current_trajectories_state_transition_pgfl}). Then the predicted distribution for the next step $G_{k|k-1}[h]$ is of the form (\ref{eq:SetCPUnion}), with:
\begin{subequations}
	\begin{align}
		\lambdau_{k|k-1}(\traj_k) &= \lambda_{k}^{\rm B}(\traj_k) + \conv{\lambdau_{k-1|k-1}}{\pi^{\rm c}P_{k-1}^{\rm S}}\label{eq:UndetectedPropagation} \\
		n_{k|k-1} &= n_{k-1|k-1} \label{eq:PredictedNumberOfTracks}\\
		h^i_{k|k-1} & = h^i_{k-1|k-1} \; \forall \; i \label{eq:PredictedAssociationWeight} \\
		w_{k|k-1}^{i,a^i} &= w_{k-1|k-1}^{i,a^i} \; \forall \; i, a^i \\
		r_{k|k-1}^{i,a^i} & = r_{k-1|k-1}^{i,a^i}\conv{f_{k-1|k-1}^{i,a^i}}{P_{k-1}^{\rm S}}, \; \forall \; i, a^i \label{eq:ExistProbPropagation} \\
		f_{k|k-1}^{i,a^i} & = \frac{\conv{f_{k-1|k-1}^{i,a^i}}{\pi^{\rm c}P_{k-1}^{\rm S}}}{\conv{f_{k-1|k-1}^{i,a^i}}{P_{k-1}^{\rm S}}}, \; \forall \; i, a^i \label{eq:KinematicDistPropagation}
	\end{align}%
\label{eq:PredictionCurrentTrajectories}%
\end{subequations}%	
\end{theorem}%
%In \eqref{eq:UndetectedPropagation} we see how the predicted set of undetected targets consists of those that survive from the previous time step, plus the ones that are born at this time. The number of tracks and the association weights are both constant through the prediction, see \eqref{eq:PredictedNumberOfTracks} and \eqref{eq:PredictedAssociationWeight}. 
%For a posterior density/intensity at time $k$ of the form \eqref{eq:mixture_trajectory_density},
%\begin{align}
%	p_{k|k}(\traj) = \sum_{j} w_{k|k}^{j} p_{k|k}^{j}(\stseq_{\tb:\td} | \tb, \td)\Delta_{\td}(e_{k|k}^{j})\Delta_{\tb}(b_{k|k}^{j}),
%\end{align}
%the following results hold:
%\begin{align}
%	\conv{p_{k|k}}{P_{k}^{\rm S}} = & \sum_{j: e_{k|k}^{j}=k} w_{k|k}^{j} \int p_{k|k}^{j}(x_{k}) P^{\rm S}(x_{k}) \diff x_{k} \\
%	\conv{p_{k|k}}{P_{k}^{\rm S}\pi^{\rm c}} = & \sum_{j: e_{k|k}^{j}=k} w_{k|k}^{j} P^{\rm S}(x_{k}) \nonumber \\
%	& \times \pi_{}^{x}(x_{k+1}|x_{k}) p_{k|k}^{j}(\stseq_{\tb:\td} | \tb,\td) \nonumber \\
%	& \times \Delta_{\td}(k+1) \Delta_{\tb}(\beta^{j})
%\end{align}
\subsubsection{Transition model for the set of all trajectories}
The Bernoulli \rfs transition density without birth is
\begin{subequations}
\begin{align}
	& f_{k|k-1}^{\rm a}(\setX | \setX') = \\
	& \left\{ \begin{array}{lcl}
			1 & & \setX'=\emptyset, \setX=\emptyset \\
			%0 & & \setX'=\emptyset, \setX\neq\emptyset \\
			%0 & \text{if} & \setX'=\{\traj'\},\setX=\emptyset \\
			\pi^{\rm a}(\traj|\traj') & \text{if} & \setX'=\{\traj'\},\setX=\{\traj\} \\
			0 & & \text{otherwise}
		\end{array} \right. \nonumber \\
	& \pi^{\rm a}(\traj | \traj') = \pi_{}^{{\rm a},\stseq}(\stseq_{\tb:\td} | \tb,\td,\traj') \pi^{\td}_{}(\td | \tb,\traj') \Delta_{\tb'}(\tb), \\
	& \pi^{\td}_{}(\td | \tb, \traj') = \left\{\begin{array}{ll} 1, & \td = {\td}' < k-1 \\ 1-P_{k-1}^{\rm S}(\traj'), & \td = {\td}' = k-1 \\ P_{k-1}^{\rm S}(\traj'), & \td = {\td}'+1 = k \\ 0, & \text{otherwise} \end{array} \right. , \\
	&\pi^{{\rm a},\stseq}_{}(\stseq_{\tb:\td} | \tb,\td,\traj') = \\
	& \left\{\begin{array}{ll} \delta_{\stseq_{\tb':\td'}'}(\stseq_{\tb:\td}), & \td = \td' \\ \pi^{x}_{}(x_{\td} | x'_{\td'}) \delta_{\stseq'_{\tb':\td'}}(\stseq_{\tb:\td-1}), & \td = \td'+1 \end{array}\right. \nonumber .
\end{align}%
\label{eq:set_of_all_trajectories_state_transition_density}%
\end{subequations}%
%and the trajectory dependent probability of survival is defined in \eqref{eq:trajectory_dependent_probability_of_survival}.
In this model, the interpretation of the probability of survival is that it governs whether or not the trajectory ends, or if it extends by one more time step. However, importantly, regardless of whether or not the trajectory ends, the trajectory remains in the set of all trajectories.

In this case, the \pgfl of the multiple object transition density with \ppp birth is 
\begin{align}
	G_{k|k-1}^{\rm a}[h | \setX] = e^{\conv{\lambda_{k}^{\rm B}}{h-1}} \conv{\pi^{\rm a}}{h}^{\setX}
	\label{eq:set_of_all_trajectories_state_transition_pgfl}
\end{align}
Note the difference to \eqref{eq:set_of_current_trajectories_state_transition_pgfl}, which comes from the way in which the probability of survival is used. In this case, the target always remains in the set of targets, even if its trajectory has ended.
%Using this second model of the time evolution of the set of trajectories, 
The prediction step is presented in the theorem below.

\begin{theorem}\label{th:PredictionAllTrajectories}
	Assume that the distribution from the previous time step $G_{k-1|k-1}[h]$ is of the form given in (\ref{eq:SetCPUnion}), and that the transition model is of the kind (\ref{eq:set_of_all_trajectories_state_transition_pgfl}). Then the predicted distribution for the next step $G_{k|k-1}[h]$ is of the form (\ref{eq:SetCPUnion}), with:
	\begin{subequations}
	\begin{align}
		\lambdau_{k|k-1}(\traj_k) &= \lambda_{k}^{\rm B}(\traj_k) + \conv{\lambdau_{k-1|k-1}}{\pi^{\rm a}}\label{eq:allUndetectedPropagation} \\
		n_{k|k-1} &= n_{k-1|k-1} \\
		h^i_{k|k-1} & = h^i_{k-1|k-1} \; \forall \; i \\
		w_{k|k-1}^{i,a^i} &= w_{k-1|k-1}^{i,a^i} \; \forall \; i, a^i \\
		r_{k|k-1}^{i,a^i} & = r_{k-1|k-1}^{i,a^i}, \; \forall \; i, a^i \label{eq:allExistProbPropagation} \\
		f_{k|k-1}^{i,a^i} & = \conv{f_{k-1|k-1}^{i,a^i}}{\pi^{\rm a}} , \; \forall \; i, a^i \label{eq:allKinematicDistPropagation}
	\end{align}%
	\label{eq:PredictionAllTrajectories}%
	\end{subequations}%		
\end{theorem}%
\vspace{-8mm}%
\subsection{Update step}

The target measurement model of assumption \ref{ass:Measurement} is extended by defining a Bernoulli measurement density as follows:
\begin{subequations}
\begin{align}
	& \varphi_{k}(\setW_{k} | \setX) = \\
	& \left\{ \begin{array}{ll}
			1, & \setX=\emptyset, \setW_k=\emptyset \\
			%0, & & \setX=\emptyset, \setW_k\neq\emptyset \\
			1-P_{k}^{\rm D}(\traj), & \setX=\{\traj\},\setW_k=\emptyset \\
			P_{k}^{\rm D}(\traj)\varphi(z_k | \traj), & \setX=\{\traj\},\setW_k=\{z_k\} \\
			0, & \text{otherwise}
		\end{array} \right. \nonumber \\
	 & P_{k}^{\rm D}(\traj) = P^{\rm D}(x_{\td})\Delta_{k}(\td), \label{eq:probability_of_detection}\\ % \left\{\begin{array}{lcl} 0 & \text{if} & \td\neq k \\ P^{\rm D}(x_{\td}) & & \td=k \end{array} \right.
	& \varphi(z | \traj) = f(z | x_{\td}).
\end{align}
\label{eq:single_object_measurement_model}%
\end{subequations}
The clutter is modelled as a \ppp with intensity $\lambda_{k}^{\rm FA}(z)$, and it follows that the measurement \pgfl is
\begin{align}
	&G_{k}[g | \setX_{k}] = e^{\conv{\lambda_{k}^{\rm FA}}{g-1}} \left(1-P^{\rm D} + P^{\rm D}\conv{ \varphi}{g}\right)^{\setX_k}. \label{eq:measurement_model_pgfl}
\end{align}
We see that $G_{k}[g | \setX_{k}] $ is a trajectory equivalent to the standard point target \pgfl, and thus the trajectory measurement update is analogous to the target measurement update in \cite{Wil12}. %
\begin{theorem}\label{th:Update}
	Assume that the predicted distribution $G_{k|k-1}[h]$ is of the form given in (\ref{eq:SetCPUnion}), and that the measurement model is of the kind (\ref{eq:measurement_model_pgfl}). Then, the updated distribution $G_{k|k}[h]$ (updated with the measurement set $Z_k=\{z_k^1,\dots,z_k^{m_k}\}$) is of the same form, with $n_{k|k} = n_{k|k-1} + m_k$, 
	{
	\begin{align}
		\lambdau_{k|k}(\traj_k) &= \left( 1-P^{\rm D}(\traj_k)\right)\lambdau_{k|k-1}(\traj_k),\label{eq:UndetectedUpdate} \\
		{\cal M}^k &= {\cal M}^{k-1} \cup \big\{(k,j)|j\in\{1,\dots,m_k\}\big\} .
	\end{align}
	}

	For tracks continuing from previous time steps ($i\in\{1,\dots,n_{k|k-1}\}$), a hypothesis is included for each combination of a hypothesis from a previous time and either a missed detection or an update using one of the $m_k$ new measurements, such that the number of hypotheses becomes $h^i_{k|k} = h^i_{k|k-1}(1+m_k)$. For missed detection hypotheses ($i\in\{1,\dots,n_{k|k-1}\}$, $a^i\in\{1,\dots,h_{k|k-1}\})$:
	\begin{subequations}
	\begin{align}
	{\cal M}^k(i,a^i) &= {\cal M}^{k-1}(i,a^i) \label{eq:MissUpdateMeasSet} \\
	w_{k|k}^{i,a^i} &= w^{i,a^i}_{k|k-1}\left(1-r_{k|k-1}^{i,a^i} \conv{f_{k|k-1}^{i,a^i}}{P^{\rm D}} \right) \label{eq:MissUpdateW} \\
	r^{i,a^i}_{k|k} &= \frac{r_{k|k-1}^{i,a^i}\conv{f_{k|k-1}^{i,a^i}}{1-P^{\rm D}}}{1-r_{k|k-1}^{i,a^i}\conv{f_{k|k-1}^{i,a^i}}{P^{\rm D}}} \label{eq:MissUpdatePex}\\
	f^{i,a^i}_{k|k}(\traj_k) &= \frac{\left(1-P^{\rm D}(\traj)\right)f_{k|k-1}^{i,a^i}(\traj)}{\conv{f_{k|k-1}^{i,a^i}}{1-P^{\rm D}}} \label{eq:MissUpdateKin}
	\end{align}
	\label{eq:BernoulliMissedUpdate}
	\end{subequations}
	For hypotheses updating existing tracks ($i\in\{1,\dots,n_{k|k-1}\}$, $a^i=\tilde{a}^i+h^i_{k|k-1} j$, $\tilde{a}^i\in\{1,\dots,h^i_{k|k-1}\}$, $j\in\{1,\dots,m_k\}$, \textit{i.e.}\xspace, the previous hypothesis $\tilde{a}^i$, updated with measurement $z_k^j$):\footnote{A hypothesis at the previous time with $r^{i,a^i}_{k|k-1}=0$ need not be updated since the posterior weight in (\ref{eq:DetUpdateW}) would be zero. For simplicity, the hypothesis numbering does not account for this exclusion.}
	\begin{subequations}
	\begin{align}
	{\cal M}^k(i,a^i) &= {\cal M}^{k-1}(i,\tilde{a}^i) \cup \{(t,j)\} \label{eq:DetUpdateMeasSet} \\
	w_{k|k}^{i,a^i} &= w^{i,\tilde{a}^i}_{k|k-1}r_{k|k-1}^{i,\tilde{a}^i} \conv{f_{k|k-1}^{i,\tilde{a}^i}}{\varphi(z_k^j|\cdot) P^{\rm D}} \label{eq:DetUpdateW}\\
	r^{i,a^i}_{k|k} &= 1 \label{eq:DetUpdatePex}\\
	f^{i,a^i}_{k|k}(\traj_k) &= \frac{\varphi(z_k^j|\traj_k)P_{k}^{\rm D}(\traj_k)f_{k|k-1}^{i,\tilde{a}^i}(\traj_k)}{\conv{f_{k|k-1}^{i,\tilde{a}^i}}{\varphi(z_k^j|\cdot) P_{k}^{\rm D}}} \label{eq:DetUpdateKin}
	\end{align}
	\label{eq:BernoulliMeasurementUpdate}
	\end{subequations}
	Finally, for new tracks, $i\in\{n_{k|k-1}+j\}$, $j\in\{1,\dots,m_k\}$ (\textit{i.e.}\xspace, the new track commencing on measurement $z_k^j$),
	\begin{subequations}
	\begin{align}
	h^i_{k|k} &= 2 \\
	{\cal M}^k(i,1) &= \emptyset, \quad
	w^{i,1}_{k|k} = 1, \quad r^{i,1}_{k|k} = 0 \label{eq:NewTargetNonExistWQ} \\
	{\cal M}^k(i,2) &= \{(t,j)\} \label{eq:PoisUpdateMeasSet} \\
	w^{i,2}_{k|k} &=  \lfa(z_k^j) + \conv{\lambdau_{k|k-1}}{\varphi(z_k^j|\cdot)P_{k}^{\rm D}} \label{eq:PoisUpdateW} \\
	r_{k|k}^{i,2} &= \frac{\conv{\lambdau_{k|k-1}}{\varphi(z_k^j|\cdot)P_{k}^{\rm D}}}{\lfa(z_k^j) + \conv{\lambdau_{k|k-1}}{\varphi(z_k^j|\cdot)P_{k}^{\rm D}}} \label{eq:PoisUpdatePex}\\
	f_{k|k}^{i,2}(\traj_k) &= \frac{\varphi(z_k^j|\traj_k)P_{k}^{\rm D}(\traj_k)\lambdau_{k|k-1}(\traj_k)}{\conv{\lambdau_{k|k-1}}{\varphi(z_k^j|\cdot)P_{k}^{\rm D}}} \label{eq:PoisUpdateKin} 
	\end{align}%
	\label{eq:new_target_update}%
	\end{subequations}%
\end{theorem}%
%For a prior density/intensity at time $k$ of the form \eqref{eq:mixture_trajectory_density},
%\begin{align}
%	& p_{k|k-1}(\traj)  \\
%	& = \sum_{j} w_{k|k-1}^{j} p_{k|k-1}^{j}(\stseq_{\tb:\td} | \tb, \td)\Delta_{\td}(e_{k|k-1}^{j})\Delta_{\tb}(b_{k|k-1}^{j}),\nonumber
%\end{align}
%the following holds,
%\begin{align}
%	&\conv{p_{k|k-1}}{P^{\rm D}\varphi}   \\
%	& = \sum_{j: e_{k|k-1}^{j}=k} w_{k|k-1}^{j} \int p_{k|k-1}^{j}(x_{k}) P^{\rm D}(x_{k}) \phi(z | x_{k}) \diff x_{k}  \nonumber\\
%	& \conv{p_{k|k-1}}{1-P^{\rm D}}  \\
%%	= & \sum_{j: e_{k|k-1}^{j}=k} w_{k|k-1}^{j} \int p_{k|k-1}^{j}(x_{k}) (1- P^{\rm D}(x_{k})) \diff x_{k}  \nonumber\\
%%	& + \sum_{j: e_{k|k-1}^{j}<k} w_{k|k-1}^{j} \\
%	& = 1 - \sum_{j: e_{k|k-1}^{j}=k} w_{k|k-1}^{j} \int p_{k|k-1}^{j}(x_{k}) P^{\rm D}(x_{k}) \diff x_{k} \nonumber
%\end{align}
\subsection{Density/intensity representation and estimation}%
\label{sec:DensityIntensityEstimation}
Note that the birth \ppp intensity \eqref{eq:PoissonBirthIntensity} is an un-normalized mixture density of the form
\begin{align}
	p_{}&(\traj) =  \sum_{j} w_{}^{j} p_{}^{j}(\stseq_{\tb:\td} | \tb, \td)\Delta_{ e_{}^{j} }( \td )\Delta_{ b_{}^{j} }( \tb ),
	\label{eq:mixture_trajectory_density}
\end{align}
where each mixture component is characterised by a weight $w_{}^{j}$, a distinct birth time $b_{}^{j}$, a distinct most recent time $e_{}^{j}$ where $b^{j} \leq e^{j}$ for all $j$, and a state sequence density $p_{}^{j}(\cdot)$. For the weights we have that $\sum_{j}w_{}^{j} = 1$ if $p_{}(\cdot)$ is a density, and $\sum_{j}w_{}^{j} \geq 0$ if $p_{}(\cdot)$ is an intensity function, e.g., a \ppp intensity. Because of the assumption that the trajectory birth intensity is an un-normalized density of the form \eqref{eq:mixture_trajectory_density}, it follows that the undetected intensity $\lambdau_{k|k'}(\traj_k)$, and all Bernoulli densities $f_{k|k'}^{i,a^i}(\traj_k)$, will be of the form \eqref{eq:mixture_trajectory_density}. This type of state density facilitates simple representations for the state sequence $\stseq_{\tb:\td}$, conditioned on $\tb$ and $\td$. %Note that alternative density/intensity representations are possible; an investigation into this is beyond the scope of this paper.

Trajectory estimation, or trajectory extraction, is the process of obtaining estimates of the set of trajectories (or set of targets) from the multi-target density. A typical approach to \mtt estimation is to select a certain global hypothesis and report estimates from it. 
%A rigorous approach to trajectory estimation is to first construct a cost function, %e.g., using a set metric like \ospa \cite{SchuhmacherVV:2008} or \gospa \cite{RahmathullahGFS:2017}, 
%and then finding the minimum cost estimate. However, many \mtt algorithm rely on estimators that are sub-optimal but easy to compute, and for which there is considerable empirical evidence that they provide accurate estimates. 
In this paper, we use a \pmbm filter estimator---e.g., any of the three discussed in \cite[Sec. VI]{GarciaFernandezWGS:2018} is valid---to identify a global hypothesis from which to extract estimates, and then extract the full trajectories directly. Note that alternative estimators are possible, representing different trade-offs between tracking accuracy and computational cost. Comparing estimators will be adressed in future work.

%An important difference to the estimation used in the \mht and the \dglmb filter is that, rather than appending the sequence of state estimates by one more state estimate, the full state sequence is estimated. A considerable benefit of this strategy is that the obtained trajectory estimates correspond to smoothed trajectories. In other words, the set of trajectories \pmbm filters can, using a simple standard estimator and a forward filter, provide smoothing estimates for the trajectories---a backwards smoothing step is not necessary. In comparison, in order for the \mht and the \dglmb filter to provide smoothed trajectories, a backwards smoothing step is necessary.

\subsection{Structure of PMBM trackers}

Two different \pmbm trackers result from the theorems: a \pmbm tracker for the set of current trajectories is given by the prediction in Theorem~\ref{th:PredictionCurrentTrajectories} and the update in Theorem~\ref{th:Update}; a \pmbm tracker for the set of all trajectories is given by the prediction in Theorem~\ref{th:PredictionAllTrajectories} and the update in Theorem~\ref{th:Update}. 
%\begin{enumerate}
%	\item A \pmbm tracker for the set of current trajectories is given by the prediction in Theorem~\ref{th:PredictionCurrentTrajectories} and the update in Theorem~\ref{th:Update}.
%	\item A \pmbm tracker for the set of all trajectories is given by the prediction in Theorem~\ref{th:PredictionAllTrajectories} and the update in Theorem~\ref{th:Update}.
%\end{enumerate}
Note that both \pmbm trackers are \trackori. For each measurement, a potential new track is initiated, see \eqref{eq:new_target_update}. As in the \pmbm filter \cite{Wil12}, for each track there is a hypothesis tree, where each hypothesis corresponds to different data association sequences for the track. %; the structure of these track hypotheses trees are the same as in the \pmbm filter \cite{Wil12}. Upon initiation, two hypotheses are created: one that corresponds to no target, i.e., the measurement was clutter; and a second one that corresponds to a target, with zero probability of existence, cf. \eqref{eq:NewTargetNonExistWQ}. 
The predictions \eqref{eq:PredictionCurrentTrajectories} and \eqref{eq:PredictionAllTrajectories} preserve the number of tracks and the number of hypotheses\footnote{The prediction \eqref{eq:allKinematicDistPropagation} results in additional mixture components \eqref{eq:mixture_trajectory_density}.}, meaning that the predictions can be implemented without approximation. In the update, additional hypotheses are created, as indicated in \eqref{eq:BernoulliMissedUpdate} and \eqref{eq:BernoulliMeasurementUpdate}.

The Bernoulli probabilities of existence $r$ have different meanings in the two trackers: for the set of current trajectories problem formulation, $r$ is the probability that the trajectory exists at the current time step $k$ and has not ended yet; in the set of all trajectories problem formulation, $r$ represents the probability that the trajectory existed at any time between $0$ and the current time step $k$.

In Theorem~\ref{th:PredictionCurrentTrajectories}, the predicted Bernoulli existence probability \eqref{eq:ExistProbPropagation} is the product of the posterior probability and $\convinline{P_k^{\rm S}}{f}$. This results in a predicted probability $r\leq1$, where equality only holds for the atypical choice $P_k^{\rm S}(x_{k})=1$. The predicted Bernoulli density \eqref{eq:KinematicDistPropagation} results in a density in which $\td=k$ with probability $=1$ and the state sequence is augmented by one more time step. This follows directly from the set of \emph{current} trajectories problem formulation: the filter does not maintain trajectories that ended before time $k$, because those are not current trajectories. Notice the important difference in Theorem~\ref{th:PredictionAllTrajectories}: here the Bernoulli existence probability is constant, and $P_k^{\rm S}(\cdot)$ enters in the integrals with $\pi^{\rm a}$. In the set of \emph{all} trajectories problem formulation, the existence probability represents existence from time $0$ to time $k$, and this is unaffected by whether or not the trajectory ends at this time or not, or if it has already ended. The probabilities of different $\td$, and the corresponding state sequences, are captured in the Bernoulli density \eqref{eq:allKinematicDistPropagation}.

In Theorem~\ref{th:Update}, the updated Bernoulli existence probability decreases following a missed detection \eqref{eq:MissUpdatePex}, unless the predicted probability is $=1$. For the set of current trajectories, the predicted probability of existence \eqref{eq:ExistProbPropagation} is typically $<1$, however, for the set of all trajectories the predicted probability of existence can be $=1$. However, remember that trajectory death ($\td<k$) is modelled in the Bernoulli densities \eqref{eq:MissUpdateKin} via the mixture components, cf. \eqref{eq:mixture_trajectory_density}. For the detection updates \eqref{eq:BernoulliMeasurementUpdate} and \eqref{eq:new_target_update}, an implication of $P^{\rm D}(\cdot)$ \eqref{eq:probability_of_detection} is that a detected trajectory must necessarily exist at time $k$. It follows that, for both problem formulations, in the updated state densities \eqref{eq:DetUpdateKin} and \eqref{eq:PoisUpdateKin} the probability of $\td=k$ is $=1$.

%%%%%%%%%%%%%%%%%%%%%%%%%%
%%%%%%%%%%%%%%%%%%%%%%%%%%
%%%%%%%%%%%%%%%%%%%%%%%%%%
%%%%%%%%%%%%%%%%%%%%%%%%%%
%%%%%%%%%%%%%%%%%%%%%%%%%%

\section{Discussion}%
\subsection{Relations to PMBM filter and MHT}%
\label{sec:filter_relations}%
Theorem \ref{th:PredictionCurrentTrajectories} is a direct re-statement of the \pmbm filter prediction, see \cite[Thm. 1]{Wil12}, to the trajectory \rfs dynamics model. Further, Theorem \ref{th:Update} is a direct re-statement of the \pmbm filter update, see \cite[Thm. 2]{Wil12}, to the trajectory \rfs measurement model. Note that regardless of which problem formulation is considered, current trajectories or all trajectories, the update is the same. An important difference between the work presented here and the work in \cite{Wil12} is that the former is for trajectory sets, whereas the latter is for target sets.

The predictions and update of the \pmbm trackers are \trackori, with one track initiated for each measurement. By comparing the track hypotheses in the two \trackori-\pmbm trackers with the track hypotheses in the \trackori-\pmbm filter \cite{Wil12}, we see that 
%We can compare the track hypotheses in the target \rfs model in \cite{Wil12}, and the two trajectory \rfs models that results from combining either the prediction in Theorem~\ref{th:PredictionCurrentTrajectories}, or the prediction in Theorem~\ref{th:PredictionAllTrajectories}, with the update in Theorem~\ref{th:Update}. 
%First, by inspection, 
the hypothesis structure is identical for the prediction and update steps. This can be seen by comparing the components of the distribution of the form \eqref{eq:SetCPUnion}, and by considering the marginalisation of the trajectories to leave only the current time step $k$. The result of the marginalisation of a trajectory Bernoulli density with probability of existence $r_{k|k'}$ and trajectory density $p_{k|k'}(\traj)$ is a target Bernoulli density, with probability of existence 
\begin{subequations}
\begin{align}
	r_{k}^{m} %& = r_{k|k'}P(\tb\leq k,\td=k) \\
	&= r_{k|k'} \sum_{\tb\leq k} \sum_{\td = k} P_{k|k'}(\tb,\td),
\end{align}
and, if $r_k^{m}>0$, state density
\begin{align}
	p_{k}^{m}(x_{k}) = \frac{1}{r_k^m} \sum_{\tb\leq k} \sum_{\td = k}   \int p_{k|k'}(x_{\tb:\td}|\tb,\td)  \diff x_{\tb:k-1} P_{k|k'}(\tb,\td).
\end{align}%
\label{eq:trajectory_marginalisation}%
\end{subequations}%
If $r^m=0$, a state density need not be defined, because the target does not exist at time $k$.

%For the trajectory representation in Section~\ref{sec:SingleTrajectoryStateRepresentation}, we define
% a marginalisation operator at the current time step $k$, denoted $\marginalisation{\cdot}{k}$, 
%as follows,
%%\begin{subequations}
%\begin{align}
%	p_{k|k'}(x_{k}) & = \marginalisation{p_{k|k'}(\traj)}{k} \\
%	& = \sum_{\tb,\td}\int p_{k|k'}(\traj)  \Delta_{k}(\td) \diff \stseq_{\tb:\td-1}.\\
%	& = \sum_{\tb,\td}\int p_{k|k'}(\stseq_{\tb:\td}|\tb,\td)P_{k|k'}(\tb,\td) \diff \stseq_{\tb:\td-1} \Delta_{k}(\td).\\
%	& = \iint p_{k|k'}(\stseq_{\tb:k}|\tb,k)P_{k|k'}(\tb,\td=k)\diff\stseq_{\tb:k-1}\diff\tb \\
%	& = \sum_{\tb}\int p_{k|k'}(\stseq_{\tb:k}|\tb,k)P_{k|k'}(\tb,\td=k)\diff\stseq_{\tb:k-1}
%	\label{eq:trajectory_marginalisation}%
%\end{align}%
%%\end{subequations}%
%Implicit in this marginalisation is the reinterpretation of dormant trajectories as being non-existent in the current time, i.e., the set of all trajectories marginalises to only contain the set of targets whose states are currently in the surveillance area. 

Regarding the general relation between target densities and trajectory densities, a result was presented in \cite[Thm. 11]{GarSve16} showing that if a trajectory \rfs distribution is marginalised to the final time to obtain a target \rfs distribution, and prediction and update steps are performed on that target \rfs distribution, then the result is equivalent to the marginalisation of the trajectory \rfs distribution \emph{after} prediction and update using the equivalent trajectory models. 

Using the marginalisation \eqref{eq:trajectory_marginalisation}, it can be shown that marginalising each predicted/updated trajectory hypothesis is the same as predicting/updating the marginalised trajectory hypothesis; the details are omitted due to page length constraints. From this, it follows that the marginalisation of the posterior \pmbm tracker density is equivalent to the \pmbm filter update of the marginalisation of the prior \pmbm tracker density. The same type of relation can be shown for the prediction.

Since the system state in the \pmbm filter \cite{Wil12} was formally an unlabelled set of target states, track continuity was not provided formally, yet a hypothesis structure similar to \mht was observed, which was exploited to provide an implicit form of track continuity. The \mht and the \pmbm filter were compared in detail in \cite{BreChi17,BreChi18}, and conditions for equivalence were identified. It follows from the \pmbm tracker/filter-relations established here that the \pmbm filter \cite{Wil12} can be understood to operate implicitly on the trajectories: it marginalises the state sequence in each prediction, and it only ever explicitly outputs the current time step. This shows the implicit track continuity in the \pmbm filter. Note also that the general similarities between the \mht and the \pmbm filter carry over to the \pmbm trackers.

\subsection{Track initiation}

The track initiation in the \pmbm trackers is of the same type as in the \pmbm filter \cite{Wil12}, as well as in the \trackori-\mht: a trajectory is defined by the measurement that initiates it. In tracking algorithms based on labelled multi-Bernoulli densities, such as the \dglmb filter \cite{VoVo13}, % and the \lmb filter \cite{ReuterVVD:2014}, 
a trajectory is defined by the labelled Bernoulli birth component.\footnote{Specifically, each Bernoulli birth component is uniquely labelled at birth, which is modelled through the prediction step; refer to \cite{VoVo13} for details.} 

One disadvantage of labelled birth is that, by its intrinsic modelling properties, it does not permit targets to be indistinguishable up until first detection. This is a significant practical drawback in cases where targets are actually indistinguishable before being detected for the first time, e.g.: when the first detection occurs a long time after target birth; or when target births are iid, which is a common modelling assumption. In fact, in the majority of cases, it is an essential character of the tracking problem that targets are indistinguishable up until first detection, after which continuity is able to be maintained (for as long as sufficient information persists). Thus, we consider it to be an important advantage that in the \pmbm tracker, the birth is modelled by a \ppp (with intensity \eqref{eq:PoissonBirthIntensity}). 

Note that after the first detection, the targets are distinguishable. From a trajectory and a given data association hypothesis, we can infer at all times the location of the target. Thus, there is no need to label the target upon initialisation.

\subsection{Estimation}
%A discussion about different estimators for the \pmbm and the \dglmb filters can be found in \cite[Sec. VI]{GarciaFernandezWGS:2018}.
%A typical approach to \mtt estimation is to select a certain global hypothesis and report estimates from it. 
The \trackori-\mht provides estimates of a target trajectory by estimating the current state at each time, and connecting estimates from different times. The \dglmb filter estimates trajectories in a similar fashion: the set of current target states is estimated at each time, and the estimates from different times are connected into trajectories using the labels. Both of these methods can be understood as answering the trajectory question \emph{``for the chosen data association hypothesis, what is the current location of the target that was initiated by measurement $z_{\tb}$ at time $\tb$ (in the case of MHT), or was initiated by birth component $l_{\tb}$ at time $\tb$ (in the case of $\delta$-GLMB)?''}

The estimation in the \pmbm tracker (Section~\ref{sec:DensityIntensityEstimation}) can be understood as answering the trajectory question \emph{``for the chosen data association hypothesis, what is the state sequence of the target that was initiated by measurement $z_{\tb}$ at time $\tb$?''} A benefit of this type of estimation is that instead of appending the sequence of estimates by one more state estimate, it estimates the full state sequence, i.e., smoothed estimates. This is possible, thanks to the fact that each global hypothesis contains full trajectory information.

Thus, the \pmbm tracker produces full trajectory estimates upon receipt of each new set of measurements, whereas the \pmbm filter, the \mht, and the \dglmb filter produce estimates of the latest time. 
%For recursive estimation, conceptually a full trajectory will be provided upon receipt of each new scan of measurements. 
The beginning of the trajectory will reveal the origin of the target, and thus correspondence between trajectories estimated at different times is established. Practically, these linkages can be made simply through hypothesis metadata, similar to \mht, and if all that is required is to estimate the current state at each time, the \pmbm filter suffices, since it may be interpreted as implicitly operating on the full trajectory (cf. Section~\ref{sec:filter_relations}).

\section{Linear Gaussian implementation}
\label{sec:linear_gaussian_implementation}

Let the single target transition density and measurement likelihoods both be linear-Gaussian,
\begin{align}
	\pi_{}^{x}(x | x') = \Npdfbig{x}{F x'}{Q}, f(z | x) = \Npdfbig{z}{H x}{R}, \label{eq:linear_gaussian_model}
\end{align}
and let the probabilities of detection and of survival be constant, $P_{k}^{\rm D}(\cdot)=P^{\rm D}$ and $P_{k}^{\rm S}(\cdot) = P^{\rm S}$.
Further, let the state sequence, conditioned on $\tb$ and $\td$, be Gaussian distributed
%\begin{align}
%	p(\stseq_{\tb:\td} | \tb,\td) = \Npdfbig{\stseq_{\tb:\td}}{\mathbf{m}}{P},
%\end{align}
with mean $m$ and covariance $P$. The state sequence density can be expressed in information form, 
%\begin{subequations}
\begin{align}
	p_{k|k'}(\stseq_{\tb:\td} | \tb,\td) %= \mathcal{N}^{-1}\left(\stseq_{\tb:\td} \ ; \ \infvec_{k|k'}, \infmat_{k|k'} \right) \\
	& = \frac{e^{-\frac{1}{2}\infvec_{k|k'}^{T}\infmat_{k|k'}^{-1}\infvec_{k|k'} -\frac{1}{2}\stseq_{\tb:\td}^{T}\infmat_{k|k'}\stseq_{\tb:\td} + \infvec_{k|k'}^{T} \stseq_{\tb:\td}}}{\sqrt{|2\pi\infmat_{k|k'}^{-1}|}} %e^{-\frac{1}{2}\stseq_{\tb:\td}^{T}\infmat_{k|k'}\stseq_{\tb:\td} + \infvec_{k|k'}^{T} \stseq_{\tb:\td}} %\\
	%= & \frac{ \exp\left(-\frac{1}{2}\left(\infmat\stseq_{\tb:\td} - \infvec\right)^{T} \infmat^{-1}\left( \infmat\stseq_{\tb:\td} - \infvec \right) \right) }{\sqrt{|2\pi\infmat^{-1}|}}
	\label{eq:information_form_gaussian}%
\end{align}%
%\end{subequations}%
with information vector $\infvec_{k|k'}=P_{k|k'}^{-1}m_{k|k'}$ and information matrix $\infmat_{k|k'}= P_{k|k'}^{-1}$.

Let $\infvec_{k|k}$ and $\infmat_{k|k}$ be the posterior information parameters. Given a linear Gaussian motion model \eqref{eq:linear_gaussian_model}, the predicted trajectory density has information vector and information matrix \cite{reustice-2006b}
\begin{subequations}
\begin{align}
	\infvec_{k+1|k} & = \begin{bmatrix} \text{\footnotesize$\infvec_{k|k}$} \\ \text{\footnotesize$\mathbb{O}_{n_x \times 1}$} \end{bmatrix} \\
	\infmat_{k+1|k} & = \begin{bmatrix} 
		\text{\footnotesize$\infmat_{k|k}^{[\tb:\td-1,\tb:\td-1]}$} & \text{\scriptsize$\infmat_{k|k}^{[\tb:\td-1,\td]}$} & \text{\footnotesize$\mathbb{O}_{(\tlen-1)n_x \times n_x}$} \\
		\text{\footnotesize$\infmat_{k|k}^{[\td,\tb:\td-1]}$} & \text{\footnotesize$\infmat_{k|k}^{[\td,\td]}+F^{T}Q^{-1}F$} & \text{\footnotesize$-F^{T}Q^{-1}$} \\
		\text{\footnotesize$\mathbb{O}_{n_x\times(\tlen-1)n_x}$} & \text{\footnotesize$-Q^{-1}F$} & \text{\footnotesize$Q^{-1}$} 
	\end{bmatrix} \label{eq:predicted_information_matrix}
\end{align}%
\label{eq:information_prediction}%
\end{subequations}%
where $n_{x}$ is the dimension of the single-time-step state $x$, $\mathbb{O}_{m\times n}$ is an $m$ by $n$ all-zero matrix, and $\infmat^{[a:b,c:d]}$ denotes the part the information matrix with rows for time steps $a$ to $b$ and columns for time steps $c$ to $d$. 
%Let $\infvec_{k+1|k}$ and $\infmat_{k+1|k}$ be the prior information parameters. 
Given a linear Gaussian measurement model \eqref{eq:linear_gaussian_model}, the posterior trajectory density has information vector and information matrix \cite{reustice-2006b}
\begin{subequations}
\begin{align}
	\infvec_{k+1|k+1} & = \infvec_{k+1|k} + \begin{bmatrix} \text{\footnotesize$\mathbb{O}_{(\tlen-1)n_x \times 1}$} \\ \text{\footnotesize$H^{T} R^{-1} z$} \end{bmatrix}, \\
	\infmat_{k+1|k+1} & = \infmat_{k+1|k} + \begin{bmatrix} \text{\footnotesize$\mathbb{O}_{(\tlen-1)n_x \times (\tlen-1)n_x}$}  & \text{\footnotesize$\mathbb{O}_{(\tlen-1)n_x \times n_x}$} \\ \text{\footnotesize$\mathbb{O}_{n_x \times (\tlen-1)n_x}$} & \text{\footnotesize$H^{T} R^{-1} H$} \end{bmatrix}. \label{eq:updated_information_matrix}
\end{align}

\end{subequations}
%\begin{align}
%	\infvec_{k+1|k+1} & = \infvec_{k+1|k} + \mathbf{H}^{T} R^{-1} z, \\
%	\infmat_{k+1|k+1} & = \infmat_{k+1|k} + \mathbf{H}^{T} R^{-1} \mathbf{H}, \\
%	\mathbf{H} & = \begin{bmatrix} \mathbb{O}_{d \times n_x} & \ldots & \mathbb{O}_{d \times n_x} & H \end{bmatrix},
%\end{align}
%see, e.g., \cite{reustice-2006b}. The update can be rewritten as
%\begin{align}
%	\infvec_{k+1|k+1} & = \begin{bmatrix} \infvec^{\tb:\td-1}_{k+1|k} \\ \infvec^{\td}_{k+1|k} + H^{T} R^{-1} z \end{bmatrix}, \\
%	\infmat_{k+1|k+1} & = \begin{bmatrix} \infmat^{\tb:\td-1,\tb:\td-1}_{k+1|k} & \infmat^{\tb:\td-1,\td}_{k+1|k} \\ \infmat^{\td,\tb:\td-1}_{k+1|k} & \infmat^{\td,\td}_{k+1|k} + H^{T} R^{-1} H \end{bmatrix}.
%\end{align}

Note that in the prediction~\eqref{eq:information_prediction}, both the information vector and the information matrix are augmented to account for the additional time step that, following the prediction, is included by the trajectory. A key result is that the bottom left and top right corners of the predicted information matrix \eqref{eq:predicted_information_matrix} are exactly zero, and that the update \eqref{eq:updated_information_matrix} only affects the part of the information matrix that is related to the current state. This means that the information matrix is exactly sparse, a direct consequence of the Markov property associated with $\pi_{}^{x}(\cdot|\cdot)$. The number of non-zero elements in the information matrix increases linearly: a trajectory from $\tb$ to $\td$ includes $\tlen = \td-\tb+1$ time steps. In the information matrix this leads to $n_{x}^{2}\left( 3\tlen -2 \right)$
%\begin{align}
%	n_{x}^{2}\left( 3\tlen -2 \right)
%\end{align}
non-zero elements. In comparison, with a Gaussian representation (with a mean vector and covariance matrix) the covariance is full and has $n_{x}^{2}\tlen^2$
%\begin{align}
%	n_{x}^{2}\tlen^2
%\end{align}
non-zero elements. Alternatively, an approximate solution has to be used, in which the trajectory density is marginalised to only contain the most recent time steps.

The \pmbm trackers were implemented using Bernoulli state densities and Poisson intensities of the form \eqref{eq:mixture_trajectory_density}, with information form densities \eqref{eq:information_form_gaussian} for the state sequences. To limit the number of data associations in each update (cf. Theorem~\ref{th:Update}), analogously to the \pmbm filter, see \cite[Sec V.C.3]{GarciaFernandezWGS:2018}, the $K$ best global hypotheses are found using Murty's algorithm~\cite{Murty:1968}. Given a Bernoulli state density, an estimate of the trajectory is obtained by selecting the most probable mixture component $j^{\star} = \arg\max_{j} w_{k|k'}^{j}$, and computing the expected value of the state sequence $\hat{\stseq}_{\hat{\tb}:\hat{\td}} = (\infmat_{k|k'}^{j^{\star}})^{-1}\infvec_{k|k'}^{j^{\star}}$, where $\hat{\tb} = b_{k|k'}^{j^{\star}}$ and $\hat{\td} = e_{k|k'}^{j^{\star}}$. Note that in theory, the weights \eqref{eq:DetUpdateW} and \eqref{eq:PoisUpdateW}, and the estimate $\hat{\stseq}_{\hat{\tb}:\hat{\td}}$, involves the inverse of the information matrix $\infmat_{k|k'}^{j^{\star}}$. However, it is not necessary to compute the inverse in practice. Instead, multiplications with the inverse information matrix are solved efficiently as a sparse least squares problem. Further discussion about computationally efficient data association, as well as state recovery (both full and partial), can be found in, e.g., \cite{reustice-2006b}. %Details are omitted here due to page length constraints.
%\begin{align}
%	j^{\star} &= \arg\max_{j} w_{k|k'}^{j}, \\
%	\hat{\tb} &= b_{k|k'}^{j^{\star}}, \ \hat{\td} = e_{k|k'}^{j^{\star}}, \ \hat{\stseq}_{\hat{\tb}:\hat{\td}} = \left(\infmat_{k|k'}^{j^{\star}}\right)^{-1}\infvec_{k|k'}^{j^{\star}}
%\end{align}

%%%%%%%%%%%%%%%%%%%%%%%%%%%%%%%%%%
%%%%%%%%%%%%%%%%%%%%%%%%%%%%%%%%%%
%%%%%%%%%%%%%%%%%%%%%%%%%%%%%%%%%%
%%%%%%%%%%%%%%%%%%%%%%%%%%%%%%%%%%
%%%%%%%%%%%%%%%%%%%%%%%%%%%%%%%%%%

%\begin{figure}
%	\includegraphics[width=0.5\columnwidth]{PMBMresults_ThreeTargetCoalescence}
%	\caption{Information form \pmbm results}
%	\label{fig:pmbm_results}
%\end{figure}

\section{Results}
In this section we present results from two simulated scenarios, one with a high number of targets and long trajectories, and one in which there is coalescence, i.e., the targets are all very close at one point. Both scenarios were generated with a 2D constant velocity motion model, see, e.g., \cite[Sec. 3]{RongLiJ:2003}, with acceleration standard deviation $\sigma_{v}$, linear position measurements with covariance $R$ The \pmbm tracker for all targets is compared to the \dglmb filter \cite{VoVo13}, in both the birth density was assumed known.

\subsection{Measuring performance}

To evaluate tracking performance, the trajectory metric $d(\cdot,\cdot)$ \cite{RahmathullahGS16a} was used with location error cut-off $c=100$, order $p=1$, and switch cost $\gamma=20$; the trajectory metric can be decomposed into a location cost, a missed target cost, a false target cost, and a switch cost. In the simulated scenarios, we apply the metric at each time step, and normalise it by the time step. This allows a comparison of how the metric evolves over time in the scenario, as opposed to only computing the metric at the final time step.

For scenarios with many targets, computing the metric becomes costly; in this case we have used the \gospa metric \cite{RahmathullahGFS:2017} instead, which decomposes into location cost, missed target cost, and false target cost. The \gospa metric is applied to the set of target states at each time step, and a trajectory metric is obtained by summing the \gospa metric over time.

\subsection{Scenario with many targets}

A scenario with $200$ time steps and $117$ true trajectories was generated and then processed by the tracker for all trajectories. Average \gospa over $100$ Monte Carlo simulations, the location cost per target per time is $0.25$, the number of missed targets per time step is $5\times10^{-4}$, and the number of false targets per time step is $0.33$. Processing the full scenario took, on averge, $154$ seconds, i.e., $0.77$ seconds per time step.\footnote{Matlab implementation on laptop with 3.1GHz processor, 16GB memory.} This shows that the \pmbm trackers are accurate in terms of tracking performance, with low location error and few cardinality errors (missed and false targets), and that they are computationally feasible even for a high number of long trajectories. 

\subsection{Scenario with coalescence}
As noted in, e.g., \cite{Wil12}, a large number of spatially separated targets is not necessarily the most difficult scenario. Therefore, we highlight a challenging scenario with three targets that start separated, come together at the mid point of the scenario, and then separate again. The parameters were $\sigma_{v}=0.5$, $R=\diag{[100,100]}$, $P^{\rm S}=0.99$, $P^{\rm D}=0.98$ and $\lambda^\mathrm{FA}=2.5\times 10^{-8}$ in the area $[-10^3,10^3]\times[-10^3,10^3]$. The birth density had a single component, centered in the surveillance area, with a covariance covering the whole surveillance area.

\begin{figure*}
	\includegraphics[width=0.2\textwidth]{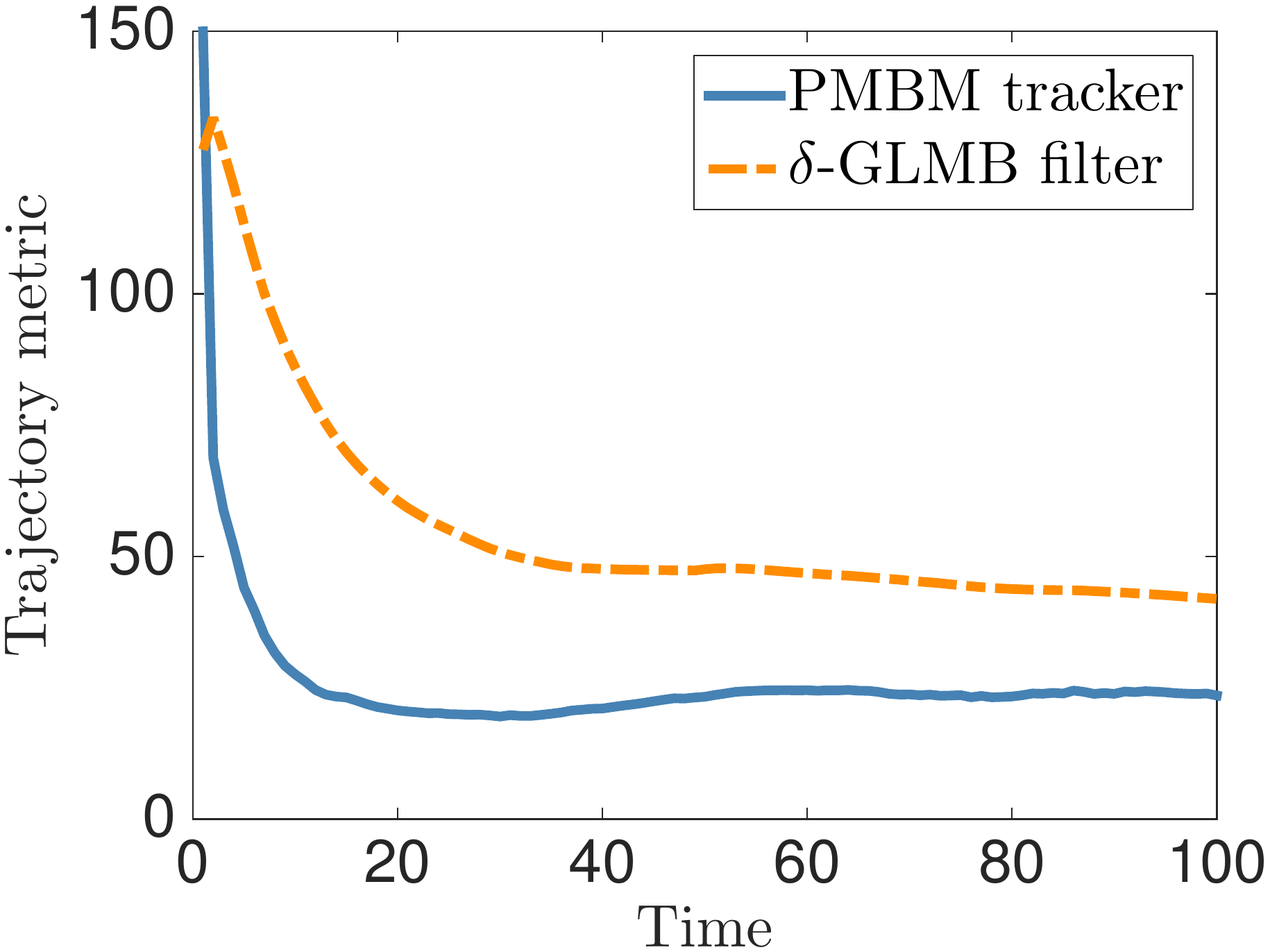}\includegraphics[width=0.2\textwidth]{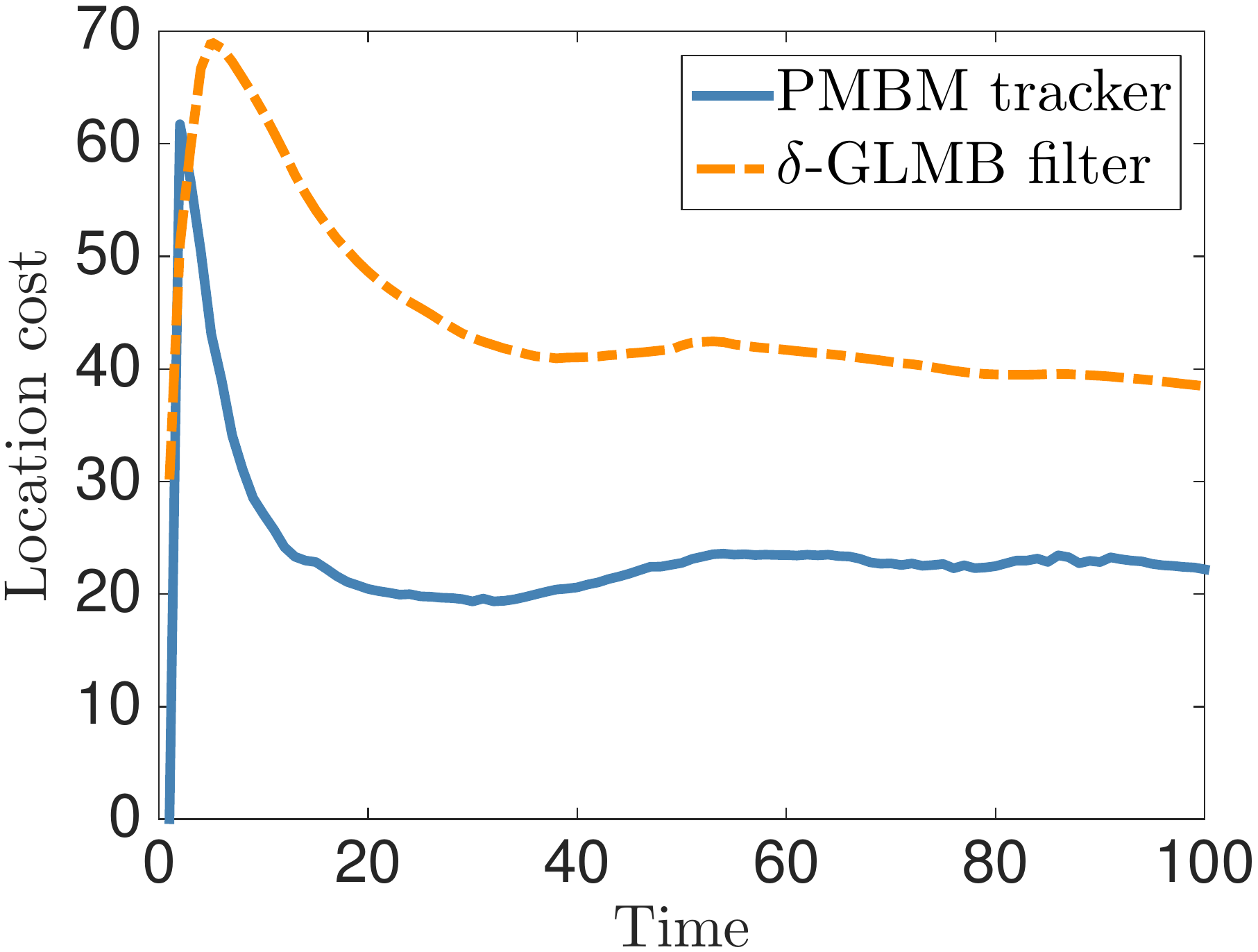}\includegraphics[width=0.2\textwidth]{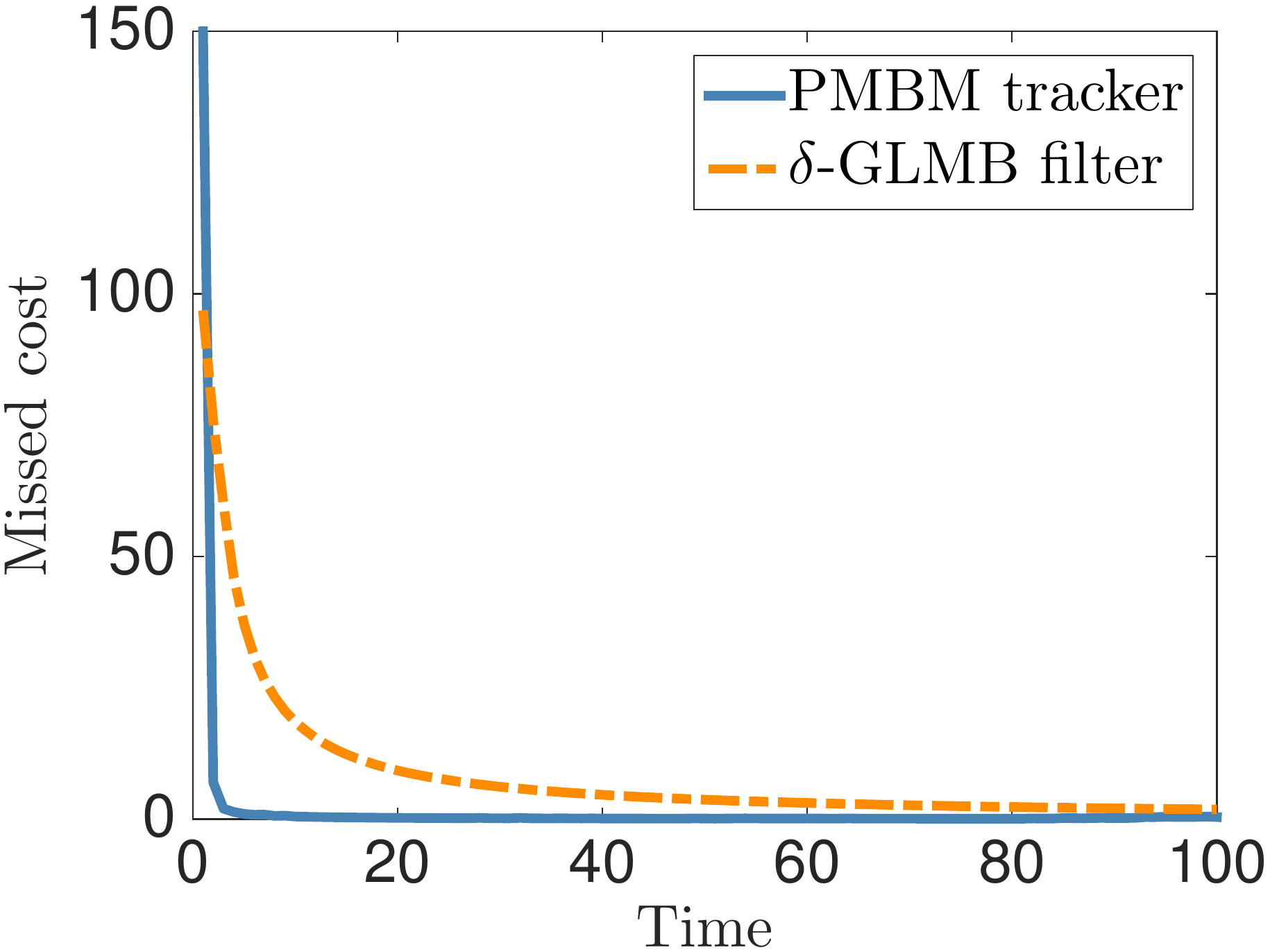}\includegraphics[width=0.2\textwidth]{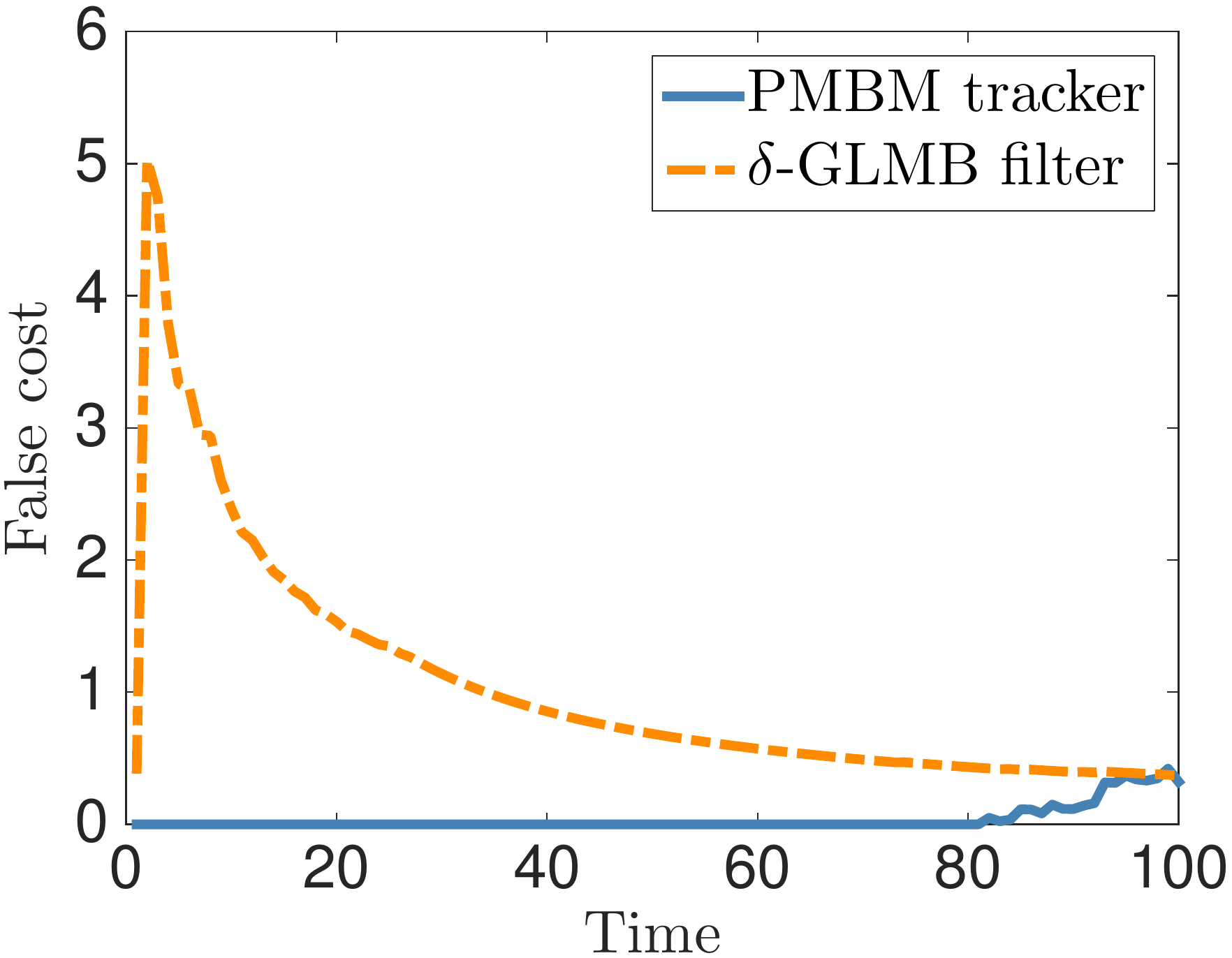}\includegraphics[width=0.2\textwidth]{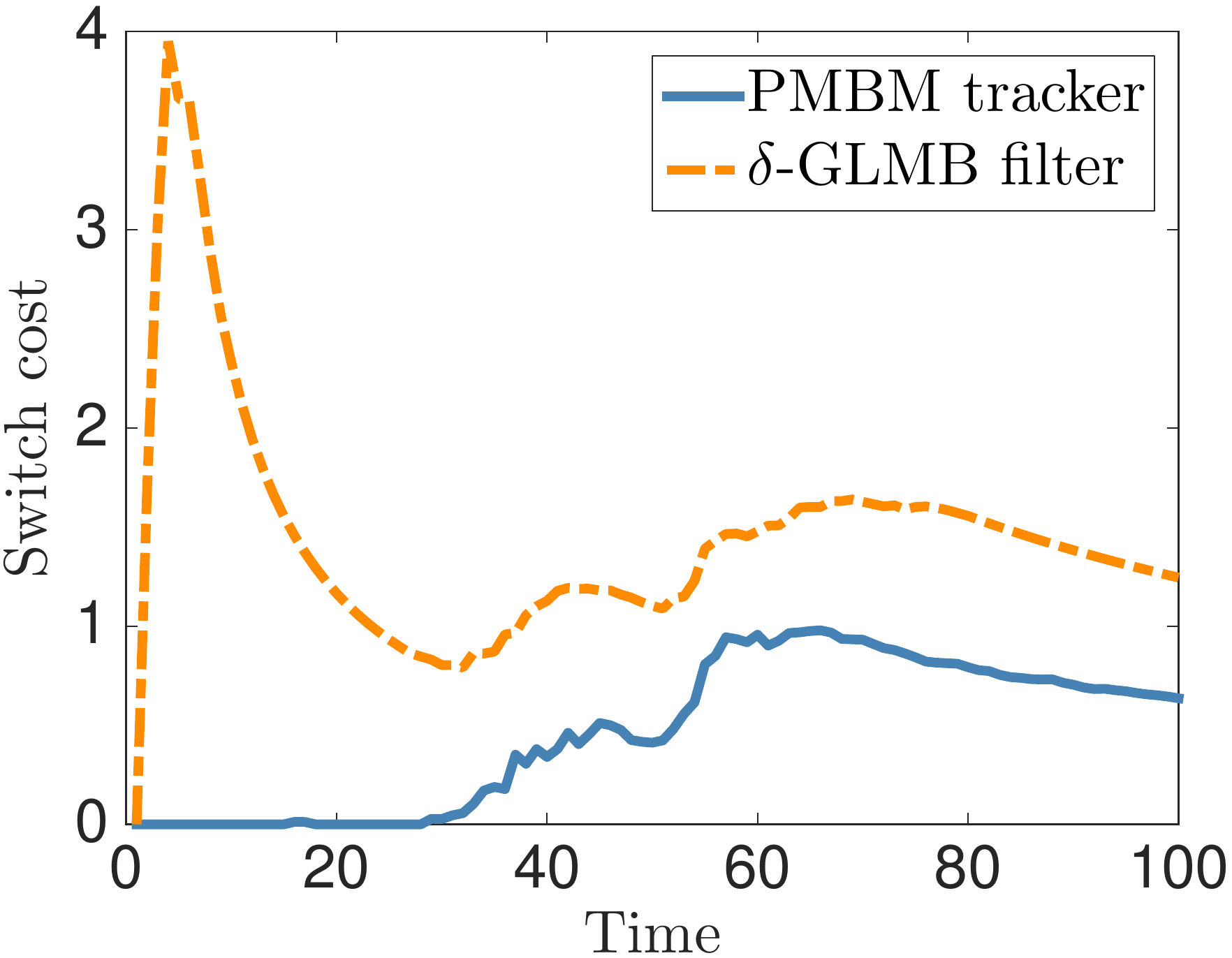}
	\caption{Results from simulated scenario with coalescence; the lines show mean over 100 Monte Carlo runs.}
	\label{fig:TrajectoryMetricResults}
\end{figure*}

The \pmbm tracker for all trajectories is compared to the \dglmb filter, which provides estimates at each time using measurements up to and including that time. For trajectory estimates extracted at the each time step of the scenario, 100 Monte Carlo simulations gave the results shown in Figure~\ref{fig:TrajectoryMetricResults}. If the results are summed over time, the following is obtained,

{\footnotesize
\begin{equation*}
	\centering
	\begin{tabular}{c | ccccc}
		      			& Metric	& Loc	& Miss	& False	& Switch  \\
		\hline
		\pmbm tracker 	& 2593.4	& 2361.7	& 181.8 	& 3.9		& 46.1 \\
		\dglmb filter     	& 5546.6	& 4442.4	& 854.3 	& 104.2 	& 145.7
	\end{tabular}%
\end{equation*}}%

\begin{figure}
	\includegraphics[width=1.0\columnwidth]{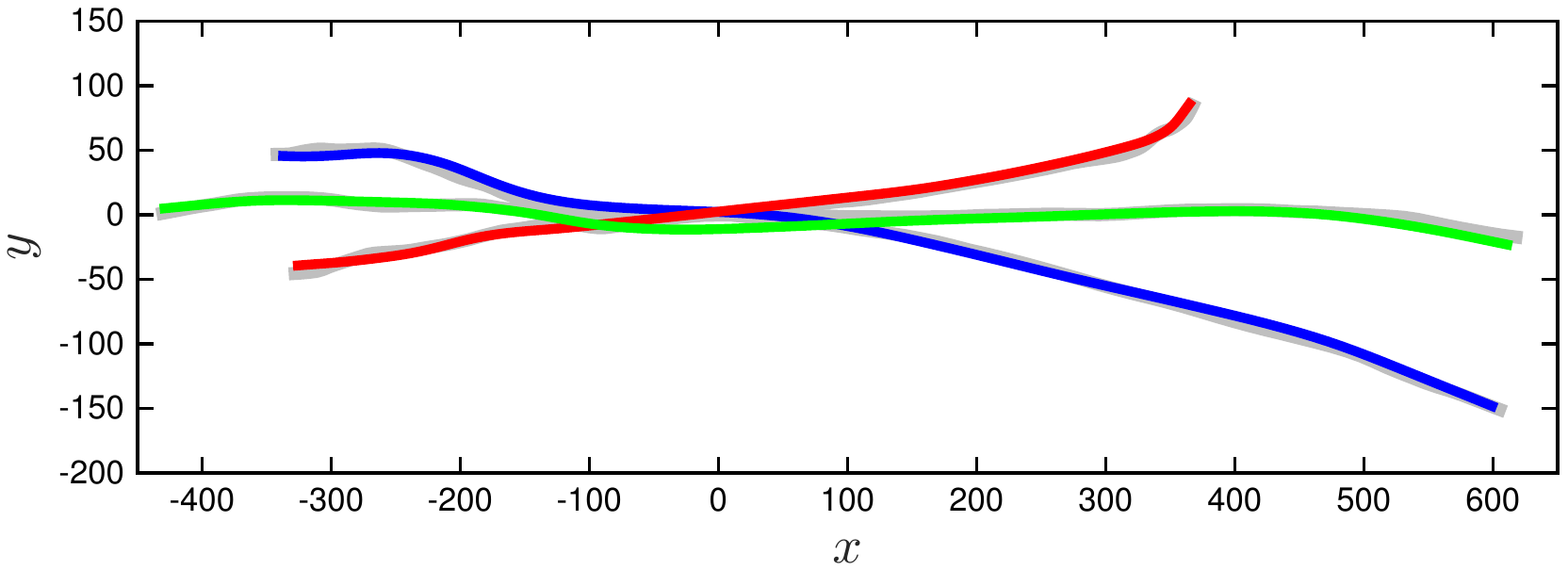}
	\includegraphics[width=1.0\columnwidth]{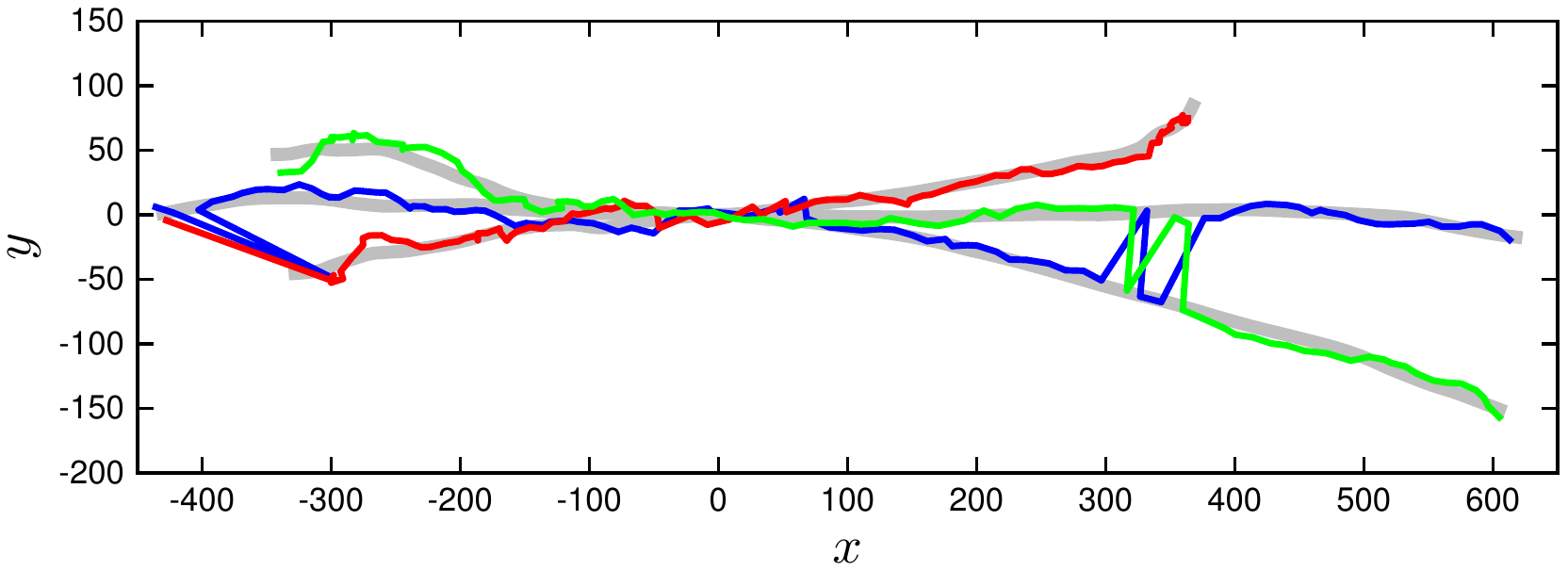}
	\caption{Example results from scenario with coalescence. Information form \pmbm results (top), Gaussian \dglmb results (bottom). Three true targets move left to right.}
	\label{fig:dglmb_results}
\end{figure}

Example results are shown in Figure~\ref{fig:dglmb_results}. The \pmbm results is three valid trajectories; however, there are switches in the \dglmb results around $(x,y)=(-350,-50)$ and $(x,y)=(300,-50)$. The switches around $(x,y)=(-350,-50)$ are due to label ambiguity upon birth: all targets are born at the same time and it is not known which birth label belongs to which measurement. The track switches around $(x,y)=(300,-50)$ are a result of ambiguity regarding which target goes where as they separate: following the separation, multiple data association sequences, i.e., global hypotheses, are approximately equally likely. Note that both the \pmbm tracker and the \dglmb filter have target switches around the mid-point of the scenario; however, it is only the \dglmb filter that sometimes produces these kinds of unrealistic switches. An important difference between the \dglmb filter and the \pmbm trackers is that, at any time, the \pmbm trackers always provide a valid trajectory, i.e., not one that is flipping between different hypotheses at different times.

\section{Conclusion}

In this paper we have presented two \pmbm trackers for the set of target trajectories, and we have established connections to the \pmbm filter for the set of targets. The connections show the implicit track continuity in the \pmbm filter.

\bibliographystyle{IEEEtran}
\bibliography{references,traj}

\end{document}